\documentclass[preprints,article,accept,pdftex,moreauthors]{mdpi} 
\firstpage{1} 
\pubvolume{1}
\issuenum{1}
\articlenumber{0}
\pubyear{2022}
\copyrightyear{2022}
\datereceived{} 
\dateaccepted{} 
\datepublished{} 
\hreflink{https://doi.org/} % If needed use \linebreak
\usepackage{subcaption}
\usepackage{soul}
\usepackage{siunitx}

\Title{Accurate and fast deep learning dose prediction for a preclinical microbeam radiation therapy study using low-statistics Monte Carlo simulations}

\TitleCitation{Accurate and fast deep learning dose prediction for a preclinical microbeam radiation therapy study using low-statistics Monte Carlo simulations}

\Author{Florian~Mentzel$^{1,*}$, Jason~Paino$^{2}$, Micah~Barnes$^{2, 5, 6}$, Matthew~Cameron$^{5}$, Stéphanie~Corde$^{2, 4, 7}$, Elette~Engels$^{2,4, 5,6}$, Kevin~Kr\"{o}ninger$^1$, Michael~Lerch$^{2, 4}$, Olaf~Nackenhorst$^1$, Anatoly~Rosenfeld$^{2, 4}$, Moeva~Tehei$^{2,4}$, Ah~Chung~Tsoi$^3$, Sarah~Vogel$^{2}$, Jens~Weingarten$^1$, Markus~Hagenbuchner$^{3,4}$ and Susanna~Guatelli$^{2, 4}$}
\AuthorNames{Florian Mentzel, Kevin Kr\"{o}ninger, Michael Lerch, Olaf Nackenhorst, Anatoly Rosenfeld, Ah Chung Tsoi, Jens Weingarten, Markus Hagenbuchner and Susanna Guatelli}

\AuthorCitation{F. Mentzel et al.}
\address{%
$^{1}$ \quad TU Dortmund University, Department of Physics, Germany\\
$^{2}$ \quad University of Wollongong, Centre for Medical Radiation Physics, New South Wales, Australia\\
$^{3}$ \quad University of Wollongong, School of Computing and Information Technology, New South Wales, Australia \\
$^{4}$ \quad Illawarra Health and Medical Research Institute, University of Wollongong, New South Wales, Australia\\
$^{5}$ \quad Imaging and Medical Beamline, ANSTO Australian Synchrotron, Victoria, Australia \\
$^{6}$ \quad Peter MacCallum Cancer Centre, Physical Sciences, Melbourne, Victoria, Australia \\
$^{7}$ \quad Prince of Wales Hospital, Randwick, New South Wales, Australia \\
}

\corres{Correspondence: florian.mentzel@tu-dortmund.de}

\simplesumm{
This work describes the development of a fast and accurate machine learning (ML) 3D U-Net dose engine, trained with Monte Carlo (MC) radiation transport simulations, to calculate the dose in rat patients treated in Microbeam Radiation Therapy (MRT) preclinical studies at the Imaging and Medical Beamline at the Australian Synchrotron. Digital phantoms are created based on CT scans of sixteen rats and are augmented to obtain enough anatomical data. Augmented variations of the digital phantoms are then used to simulate with Geant4 the energy depositions of an MRT beam inside the phantoms with 15\% (high-noise) and 2\% (low-noise) statistical uncertainty. The high-noise MC simulations are used for ML model training and validation, while the low-noise ones for testing.
The results show that the ML dose engine provides a satisfactory dose description in the tumor target and generates the dose maps in less than one second. 
} % Simple summary

\abstract{Microbeam radiation therapy (MRT) utilizes coplanar synchrotron radiation beamlets and is a proposed treatment approach for several tumour diagnoses that currently have poor clinical treatment outcomes, such as gliosarcomas. Monte Carlo (MC) simulations are one of the most used methods at the Imaging and Medical Beamline, Australian Synchrotron to calculate the dose in MRT preclinical studies. The steep dose gradients associated with the 50$\,\mu$m wide coplanar beamlets present a significant challenge for precise MC simulation of the dose deposition of an MRT irradiation treatment field in a short time frame. The long computation times inhibits the ability to perform dose optimisation in treatment planning or apply online image-adaptive radiotherapy techniques to MRT. Much research has been conducted on fast dose estimation methods for clinically available treatments. However, such methods, including GPU Monte Carlo implementations and machine learning (ML) models, are unavailable for novel and emerging cancer radiotherapy options like MRT. In this work, the successful application of a fast and accurate ML dose prediction model for a preclinical MRT rodent study is presented for the first time. The ML model predicts the peak doses in the path of the microbeams and the valley doses between them, delivered to the tumor target in rat patients. A CT imaging dataset is used to generate digital phantoms for each patient. Augmented variations of the digital phantoms are used to simulate with Geant4 the energy depositions of an MRT beam inside the phantoms with 15\% (high-noise) and 2\% (low-noise) statistical uncertainty. The high-noise MC simulation data are used to train the ML model to predict the energy depositions in the digital phantoms. The low-noise MC simulations data are used to test the predictive power of the ML model. The predictions of the ML model show an agreement within 3\% with low-noise MC simulations for at least 77.6\% 
of all predicted voxels (at least 95.9\% of voxels containing tumour) in the case of the valley dose 
prediction and for at least 93.9\% of all predicted voxels (100.0\% of voxels containing tumour) in the case 
of the peak dose prediction. The successful use of high-noise MC simulations for the training, which are much faster to produce, accelerates the production of the training data of the ML model, and encourages transfer of the ML model to different treatment modalities for other future applications in novel radiation cancer therapies.}


\begin{document}

\section{Introduction}
\noindent In recent years, an increasing number of studies investigating fast dose predictions for radiotherapy treatment planning with GPU algorithms \cite{Schiavi2017} and deep learning models has been published \cite{Kontaxis2020, Pastor-Serrano2021, Jensen2021, Mentzel2022a}. However, those publications mostly focus on clinically available treatment methods such as IMRT \cite{Brahme1982, Huang2021}, VMAT \cite{Otto2008, Lempart2021} or proton pencil beam scanning \cite{Pastor-Serrano2022}. This results partly from the urge for fast dose prediction models for those to improve clinical treatment plan optimisation capabilities \cite{Lee2022}, but also from the large available datasets from hospitals and their already delivered treatments (e. g. \cite{Torrente2022}), facilitating the development of machine learning (ML) models. Such large databases are difficult to obtain for novel and preclinical treatments. 

% \noindent This study presents an accelerated development process suited for preclinical treatments. The described workflow is developed and tested in a retrospective treatment planning scenario for microbeam radiation therapy (MRT) \cite{Slatkin1995}, resulting in the first adaption of ML models to a close-to-realistic MRT treatment scenario. In addition to small available training data, the spatially fractionated nature of this preclinical treatment exhibits additional computational complexity concerning dose estimation. MRT utilizes arrays of coplanar, approximately 50$\,\mu$m wide photon beams to achieve high \textit{peak doses} in the path of those beams with comparatively low \textit{valley doses} in between them \cite{Bartzsch2020}. Several preclinical studies have shown potential treatment benefits of MRT for tumours with poor treatment outcomes, such as radioresistant melanoma \cite{Potez2019} and gliosarcoma \cite{Engels2020}. While this work is focused on the application of the developed machine learning (ML) model and its workflow, articles on the working principles and recent progress of MRT are available in the literature \cite{Slatkin1995, Brauer-Krisch2015, Bartzsch2020}.

\noindent This study presents an accelerated development process suitable for preclinical treatments where available training data is limited. Here, our development process is applied to a novel treatment technique, Microbeam Radiation Therapy (MRT) \cite{Slatkin1995}, which presents several additional challenges concerning dose calculation. In addition to the limited available training data, the generation of Monte Carlo (MC) data sets is computationally expensive due to the high statistics required to calculate the dose depositions with few percent of statistical uncertainty, resulting from the 24-100 $\mu m$ wide microbeams, spatially fractionated with a pitch of 100-400 $\mu m$. These spatially fractionated beams result in high \textit{peak dose} regions with comparatively low \textit{valley doses} in between \cite{Bartzsch2020}. Several preclinical studies have shown potential treatment benefits of MRT for tumours with poor treatment outcomes, such as radioresistant melanoma \cite{Potez2019}, gliosarcoma and lung carcinoma~\cite{Serduc2009, Bouchet2016, Engels2020, Trappetti2021}. It is also understood that maximizing the peak-to-valley dose ratio (PVDR) results in better biological outcomes \cite{Smyth2019}, thus, accurate estimation of both peak and valley doses is necessary. While this work is focused on the application of the developed ML model and its workflow, articles on the working principles and recent progress of MRT are available in the literature~\cite{Slatkin1995, Brauer-Krisch2015, Bartzsch2020}.

% \noindent Recent proof-of-concept studies deploying ML models to estimate the doses for MRT \cite{Mentzel2022, Mentzel2022Proceedings} rely on Monte Carlo (MC) simulations with software tools like Geant4 \cite{Agostinelli2003} to create datasets to train the ML models. Even the fastest existing dose calculation methods for MRT \cite{Donzelli2018} require approximately half an hour for one prediction, hindering effective plan optimisation. 
\noindent For clinically available treatments, delivered treatment plans and computed dose distributions in previous patients can often be used as training data, however, such data does not exist for novel treatments and preclinical studies. Instead, frequently evolving phantom designs and irradiation scenarios require new training data for ML models to be calculated on a semi-regular basis. This renders the development of ML models unviable for many novel treatments. 

\noindent Recent proof-of-concept studies deploying ML models to estimate the doses for MRT \cite{Mentzel2022, Mentzel2022Proceedings} rely on MC simulations with software tools like Geant4 \cite{Agostinelli2003} to create datasets to train the ML models. Even the fastest existing dose calculation methods for MRT \cite{Donzelli2018} require approximately half an hour for one prediction with adequate statistics, hindering its effective use in treatment plan optimisation.

% \noindent For clinically available treatments, already delivered treatment plans and computed dose distributions in previous patients can often be used as datasets, which is not possible for novel treatments and preclinical studies. Instead, frequently changing phantoms and irradiation scenarios make the creation of a whole new dataset from scratch a common requirement. This renders the development of ML models less viable for many novel treatments. 

\noindent This study shows that dose estimations with satisfactory accuracy can be obtained within milliseconds with the developed ML model even with a small amount of training data available by implementing data augmentation techniques and training the models with relatively low statistics (15\% noise) MC data which are significantly faster to acquire. 

\noindent The rest of this paper is structured as follows. Section~\ref{Methods} describes the setup of the used MC simulation to generate the dose distribution data. This is followed by a description of the rat CT scans and digital phantoms used in this study. Then, the method to train and test the ML dose engine is presented. Section~\ref{results} presents the model optimisation results and the application to realistic test patient data. Finally, the findings are discussed and summarised in Section~\ref{discussion} and~\ref{conclusion}.

\section{Materials and Methods\label{Methods}}
\subsection{MC simulation\label{MC}}
\noindent In this work, an existing Geant4 simulation~\cite{Dipuglia2019, Paino} was adopted to model the generation and transport of synchrotron radiation at the Australian Synchrotron's Imaging and Medical Beamline~\cite{Stevenson2017}. The position, energy and direction of each photon of individual microbeams was recorded in a Phase Space File (PSF) just before entering the target. Then, in the Geant4 simulation developed and used in this work, the PSF is used to describe the incident radiation field on the target and to calculate the associated energy deposition in the treatment target. The advantage of this approach is the ability to use the same PSF to calculate the dose in different targets/anatomies, speeding up the overall MC simulation executions in preclinical research for MRT (e. g.~\cite{Engels2020}). The simulated microbeam field is rectangular and has a fixed size of 8x8$\,$mm$^2$, adopted from the applied treatment protocol of the preclinical study this work is based on ~\cite{Engels2020, Paino}. 

\noindent The Geant4 pre-built electromagnetic physics constructor \textit{EmStandardPhysics Option 4}~\cite{Arce2021} is adopted to model the interactions of photons and electrons in the simulation geometry. The effect of polarisation on photons processes is  considered by using the Livermore models~\cite{Geant4PhysicsManual}. All MC simulations are performed using Geant4 10.6p02~\cite{Agostinelli2003}. 

\subsection{Energy deposition scoring in voxels implemented in the MC simulation \label{scoring}}
\begin{figure}[t]
	\hspace*{-20mm}
	\centerline{
		\begin{subfigure}[t]{0.6\textwidth}
			\includegraphics[width=\linewidth]{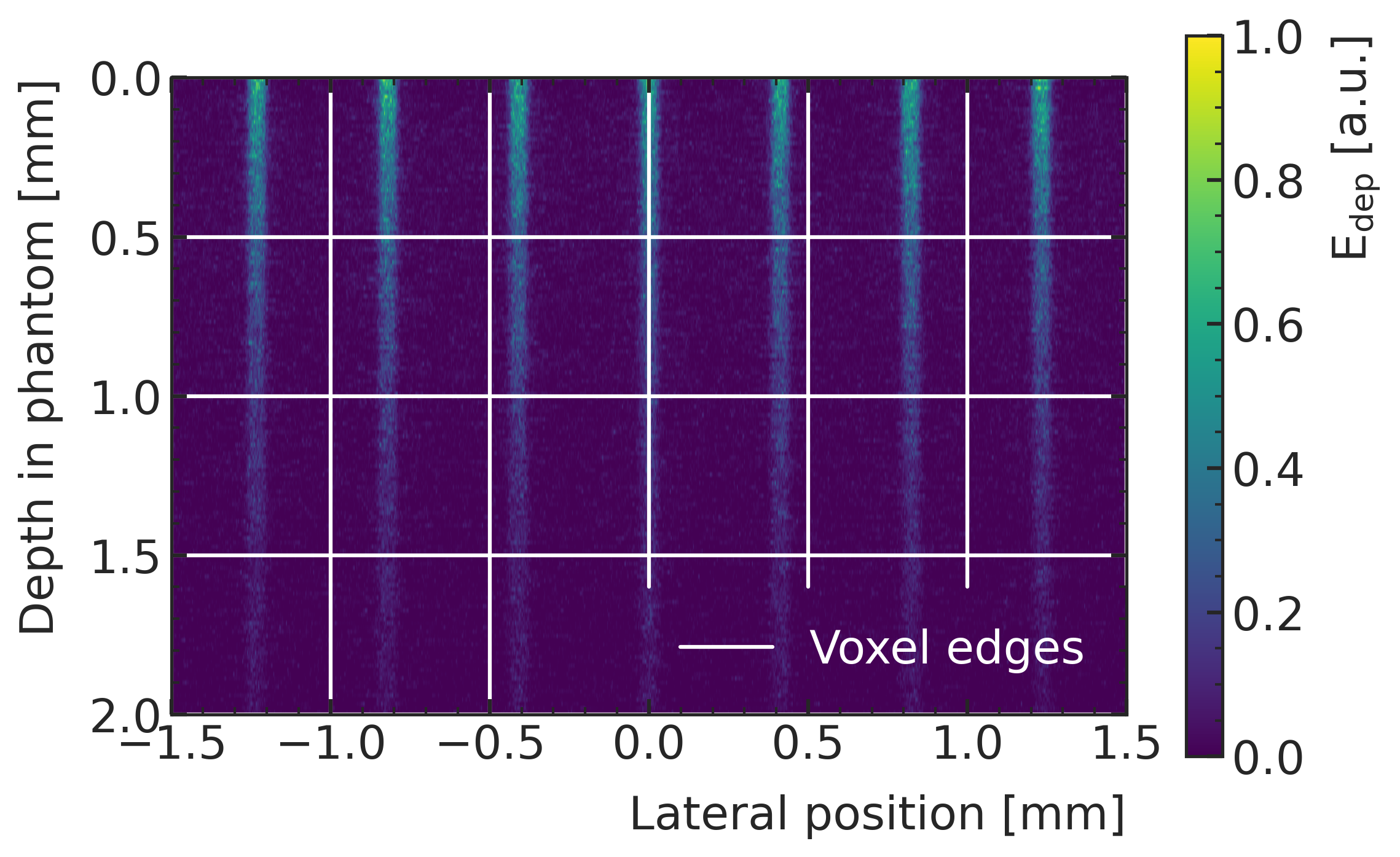}
			\caption{}
			\label{fig:MRTDose}
		\end{subfigure}
		\begin{subfigure}[t]{0.75\textwidth}
			\includegraphics[width=\linewidth]{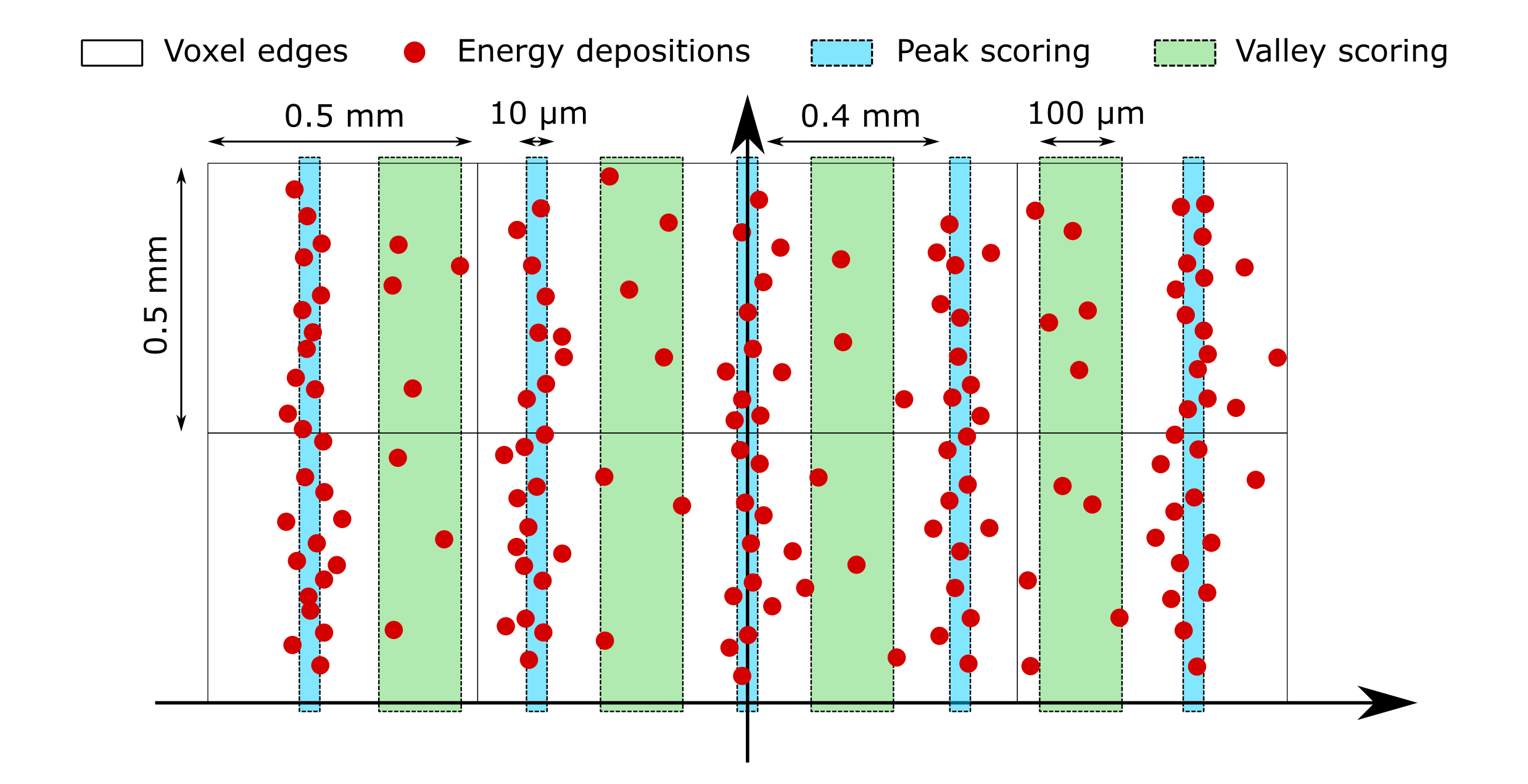}
			\caption{}	\label{fig:methods:hits_and_peakValley}
		\end{subfigure}
	}
	\caption{(a) Energy deposition in water by a sub-set of co-planar X-ray microbeams, typical of MRT, entering from the top of the shown region. The dose is deposited mainly along the tracks of the X-rays (peaks regions). The peaks are separated by valleys where the dose is significantly lower. Macro voxels are shown in white. (b) Sketch showing the concept of scoring energy deposition into macro voxels (represented as white pixels with 0.5 mm lateral sizes). The energy deposition is calculated in the peaks (light blue regions) and in the valleys (green regions) and then associated to the \textit{macro} voxel containing it. Adapted from \cite{Mentzel2022Proceedings}.}\vspace{.2cm}
\end{figure}	

Figure~(\ref{fig:MRTDose}) shows the energy deposition produced by the central 3 $mm$ wide region of the microbeam field entering a water phantom. The dose is deposited mainly along the tracks of the X-rays (peaks regions). The peaks have a width of few micrometers and are separated by valleys where the dose is significantly lower. In this study the pitch between two peak is 400~$\mu m$. MC simulations developed for MRT dosimetry usually adopt a micrometer-sized voxelisation ~\cite{Dipuglia2019, Paino} to describe the dose with satisfactory spatial resolution, however this method is computationally not feasible in the scope of this study. Instead, in this work, the energy depositions in the pathway of the peaks (width of the scoring window: 10$\,\mu$m), and the energy depositions in the valleys between them (width of the scoring window: 100$\,\mu$m), are scored separately and then assigned to the respective macro voxels (represented as white pixels in Figure~ (\ref{fig:methods:hits_and_peakValley})). This approach allows a macroscopic description of peak and valley doses and is more feasible for ML predictions as it significantly reduces the number of required voxels in the prediction volume. The PVDR is calculated as ratio between the dose in the peak and in the valley for each macro voxel. A volume of $48\times8\times8$~$mm^3$ (depth x width x width) is recorded using this technique with a macro voxel size of $0.5 \times 0.5 \times 0.5$~$mm^3$, resulting in $96 \times 16 \times 16$ voxels for each data sample.

\noindent For each geometrical configuration (corresponding to an individual rat head phantom), the simulation has been repeated twenty times with different random seeds. Then, the results of the repeated simulations have been used to calculate the mean value and standard error of the energy deposition in each voxel.   
\subsection{Rat head phantoms\label{phantoms}}
\noindent The digital phantoms of the rat heads used in this work are based on CT scans of a total of 16 rats, two weeks after implanting 9L gliosarcoma cells~\cite{Chung1983}  sourced from the European Collection of Cell Cultures (ECCC). The age of the rats is approximately six weeks at cell implantation. The rats are imaged and treated  eleven and twelve days post implantation, respectively. The average body weight is 184.5$\pm$ 9.2~g on treatment day~\cite{Engels2020}. The CT Scanner has a pixel spacing of 0.4-0.6$\,$mm and a slice thickness of 0.6$\,$mm. 

\begin{figure}[t]
	\centering 
	\includegraphics[width=\linewidth]{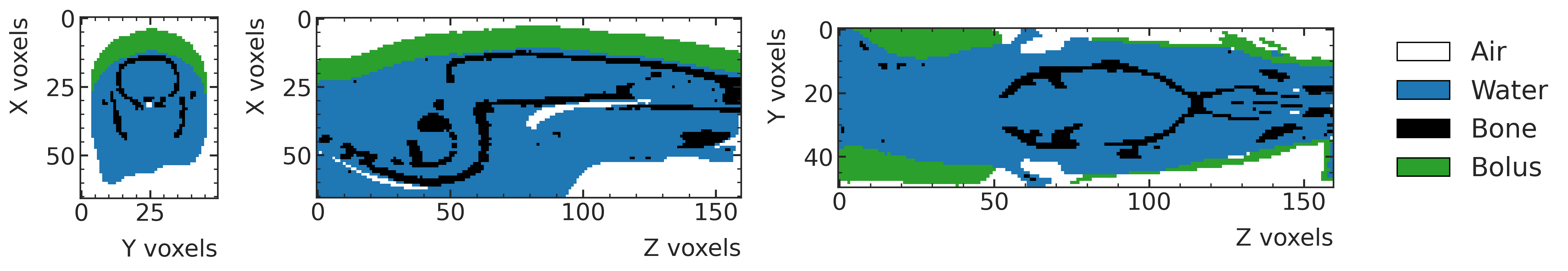}
	\caption{Example of digital rat phantom obtained from the segmentation of CT scans. Defined materials (air, water and bone) are assigned to individual voxels. Green voxels are associated to the bolus, modelled as water.}
	\label{fig:methods:rat_cts}
\end{figure}

\noindent The CT scans are used to create rats digital phantoms which are then imported into the MC simulation following the workflow detailed in~\cite{Paino}. In the first step, the centres of the brain of all CTs are manually identified and the CTs are rotated resulting in a skull orientation perpendicular to the X-axis which coincides with the beam direction. The Hounsfield Units (HU) from the CT scan are used to manually categorize the phantom voxels into three material classes: air (G4\_AIR~\cite{geant4materials}), water (G4\_WATER~\cite{geant4materials}) and bone (G4\_BONE\_COMPACT\_ICRU~\cite{geant4materials}). Finally, a 5$\,$mm thick bolus layer (G4\_WATER~\cite{geant4materials}) is placed on top of the rat phantom as per experimental setup. An example of a digitized rat phantom is shown in Figure~(\ref{fig:methods:rat_cts}). More details on the segmentation process are provided in~\cite{Paino}.
%\begin{figure}[t]
	%\centering
	%\includegraphics[width=.9\linewidth]{figure_1_1.png}
	%\caption{Three orthogonal view planes of an exemplary CT scan (rat Nr: 1).}
	%	\label{fig:methods:preprocessed_ct}
	%\end{figure}

%\begin{figure}[t]
	%\centering
	%\includegraphics[width=.75\linewidth]{figure_2_1.png}
	%\caption{Exemplary indications of the centre of phantom rotations (red cross) and the ranges of phantom translations along the three axes (blue lines indicating limits).\vspace{.5cm}}
	%\label{fig:methods:limit_indications}
	%\end{figure}

\subsection{High-noise Monte Carlo simulation datasets for training and validation \label{highnoisedata}}

\noindent 
Given the limitation that only the CT scans of sixteen rats were available for this study, we use ten scans for the training data to obtain the largest possible variety, and three scans each for the validation and test data to obtain a statistically meaningful variation for the performance evaluation during the hyperparameter optimization and for the final unbiased assessment. Rat CT scans were in no particular order in the dataset. Therefore, selecting the first ten for training is equivalent to a random selection of 10 samples.

\noindent To maximize the available training data from the limited patient CT data, data augmentation is performed to increase the number of samples by artificially generating samples. This is achieved by randomly applying transformations to the digital phantoms before running the MC simulations: translating them ($\pm$~5$\,$mm up and down from the beam's view, perpendicular to the beam as far as the beam still targets the brain), rotating around the centre of the brain ( $\pm$~10 degrees around each axis) and scaling the size of voxels isotropically in the three dimensions (factor 0.8-1.2). With this method, a total of 6500 simulation data samples are produced: 4569 samples (generated with rats number 1-10) for training, 1431 samples (generated with rats number 11-13) for validation and 500 samples (generated with rats number 14-16) for testing. 

\begin{figure}[t]
	\centering
	\hfill
	\begin{subfigure}[t]{0.4\textwidth}
		\includegraphics[width=\linewidth]{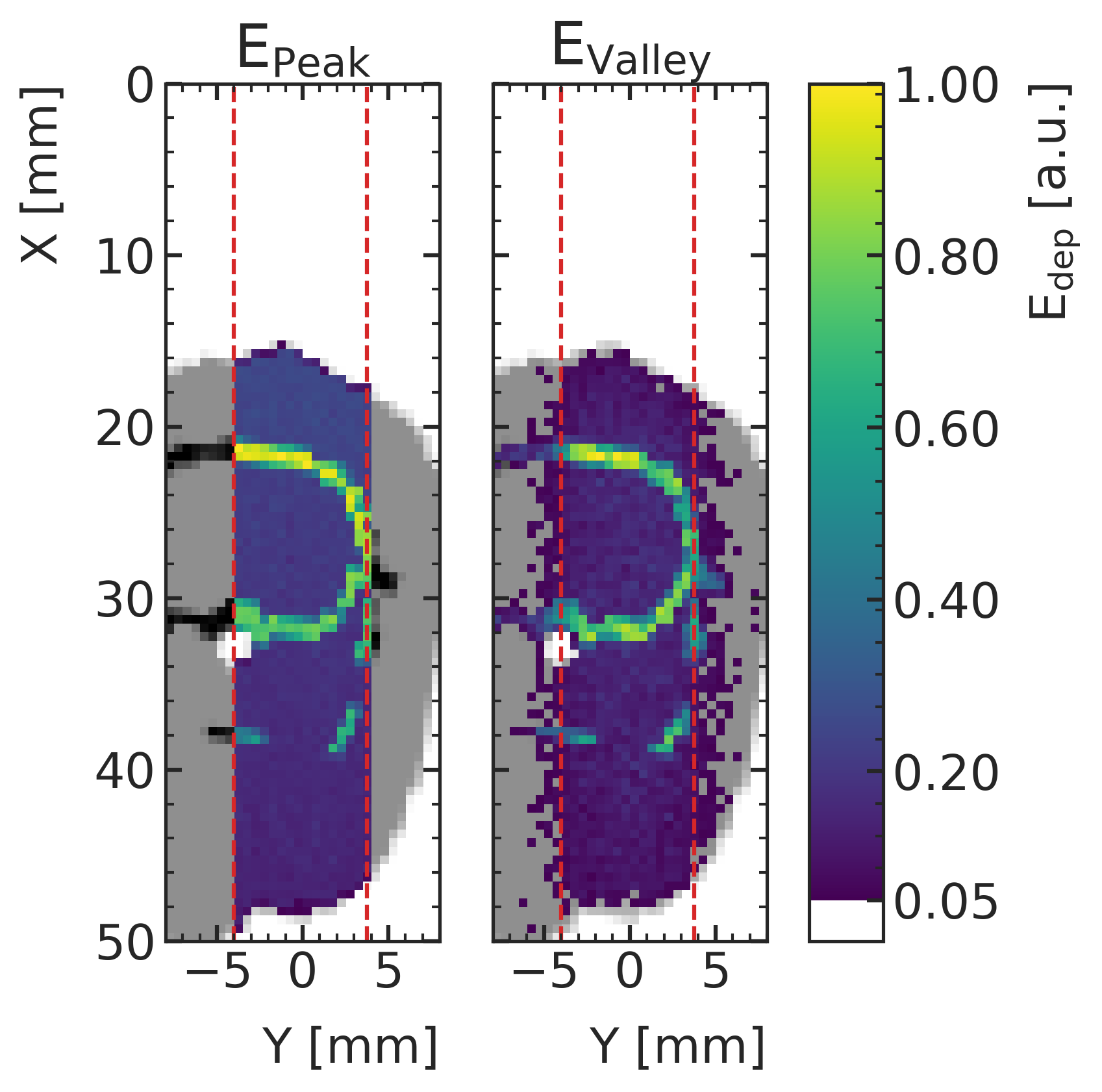}
		\caption{}
		\label{fig:methods:test_rats_2D_dose:a}
	\end{subfigure}
	\hfill
	\begin{subfigure}[t]{0.53\textwidth}
		\includegraphics[width=\linewidth]{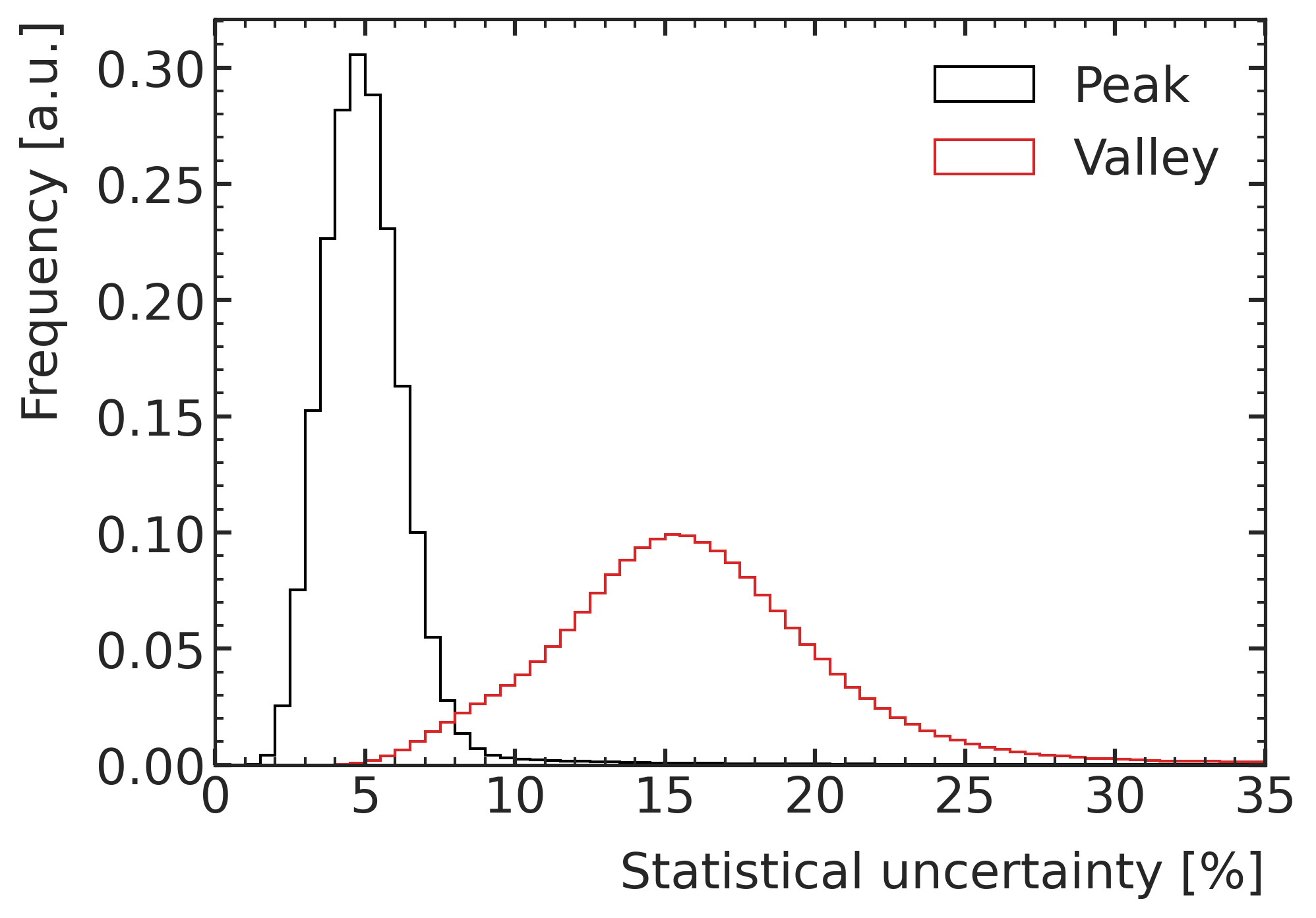}
		\caption{}
		\label{fig:methods:test_rats_2D_dose:b}
	\end{subfigure}
	\hfill
	\caption{(a) 2D slice of the MC-simulated energy deposition in the peak (left) and valley (right) respectively at the centre of the prediction volume for an exemplary high-noise training sample (rat number 1), normalized to their respective maximum. The prediction volume is indicated with red dashed lines. Air is shown white, tissue (water) in grey and bone in black. (b) Histograms of the voxel-wise statistical uncertainties (quantified with the standard error) of the peak and valley energy deposition MC simulations in the high-noise datasets.}
	\label{fig:methods:test_rats_2D_dose}
\end{figure}

\noindent Figure~(\ref{fig:methods:test_rats_2D_dose:a}) shows an example of an energy deposition map in a peak and a valley, in the central plane of the digital phantoms of the training data. The distribution of statistical uncertainty of the voxels, quantified with the standard error, peaks around 15\% for the valleys and around 5\% for the peaks, as it can be seen in ~Figure~(\ref{fig:methods:test_rats_2D_dose:b}). MC simulations with this type of uncertainty are referred as \textit{high-noise} in this paper and are used for ML training and validation. 

\subsection{Low-noise Monte Carlo simulation datasets for testing\label{Tumours}}

\begin{figure}[t]
	\centering
	\includegraphics[width=.9\linewidth]{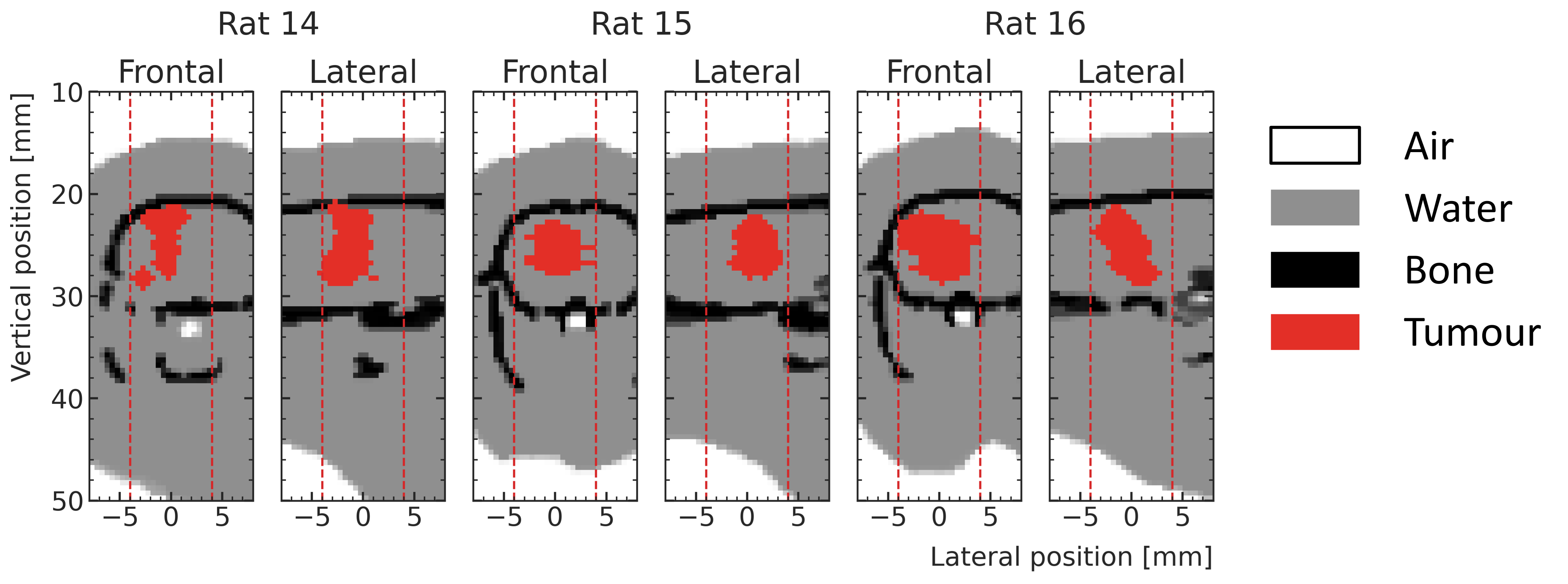}
	\caption{Frontal and lateral slices at the centre of the ML prediction regions (red dotted lines) showing exemplary tumours (red) in the respective phantoms (white - air, grey - water, black - bone) of the three test rats, used in the testing.}
	\label{fig:methods:realistic_tumours}
\end{figure}
\noindent In addition to the high-noise test samples, three \textit{low-noise} test samples are simulated at the actual tumour positions (shown in Figure~(\ref{fig:methods:realistic_tumours})) for the test rats number 14, 15 and 16. These samples are used to compare the dose predictions of the ML model with the MC simulations in the whole prediction region but with special attention to the tumour volume, for realistic, delivered treatments, without being dominated by statistical noise. %The tumours are manually segmented in the CTs %after injecting an iodine contrast enhancement %agent. The contrast agent is not present during %MRT irradiation. Two slices (frontal and %lateral) through the tumour at the centre of %the respective treatment fields are shown in %red on top of the CT scans in Figure~(\ref{fig:methods:realistic_tumours}). 

\begin{figure}[t]
	\centering
	\hfill
	\begin{subfigure}[t]{0.4\textwidth}
		\includegraphics[width=\linewidth]{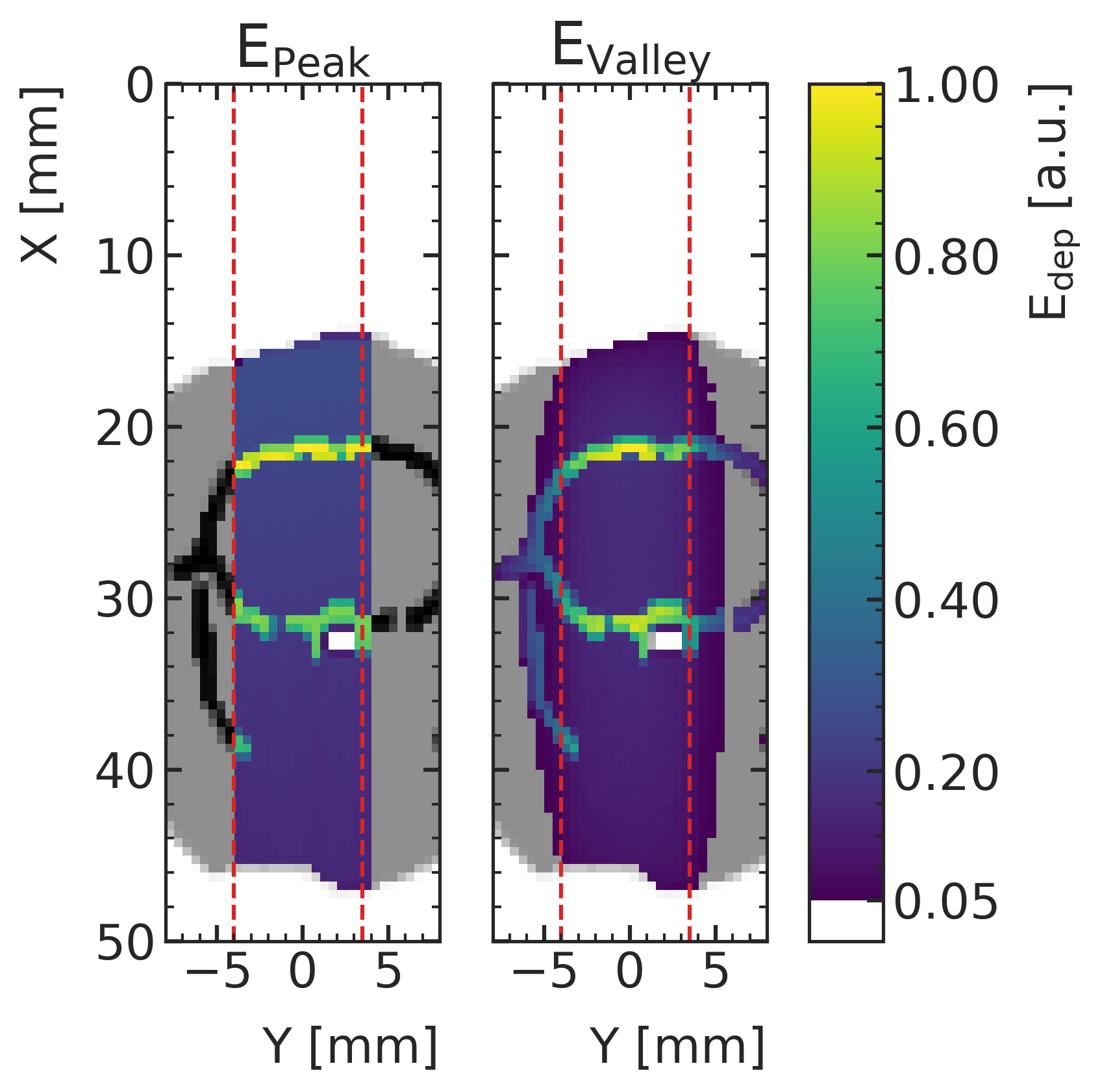}
		\caption{}
		\label{fig:stat_unc:a}
	\end{subfigure}
	\hfill
	\begin{subfigure}[t]{0.53\textwidth}
		\includegraphics[width=\linewidth]{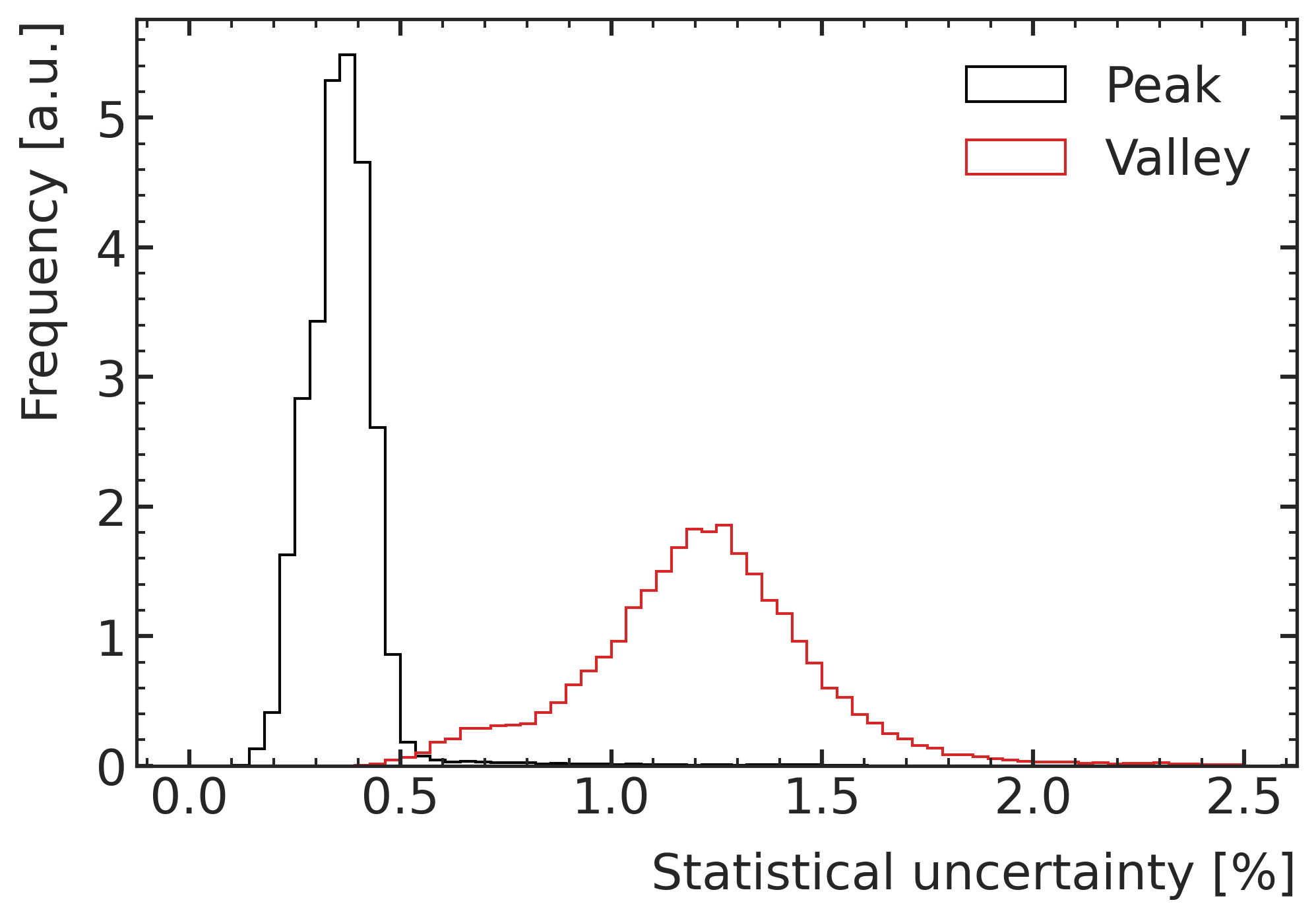}
		\caption{}
		\label{fig:stat_unc:b}
	\end{subfigure}
	\hfill
	\caption{(a) 2D slice of the MC-simulated energy deposition in the peak (left) and valley (right) respectively at the centre of the prediction volume for an exemplary low-noise training sample (rat number 15), normalized to their respective maximum. Air is shown white, tissue (water) in grey and bone in black. (b) Histograms of the voxel-wise statistical uncertainties (quantified with the standard error) of the peak and valley energy deposition MC simulations of the three low-noise treatment test data samples.}
	\label{fig:stat_unc}
\end{figure}

\noindent The statistical uncertainties compared to the high-noise datasets are significantly lower in these treatment test samples. Figure~(\ref{fig:stat_unc:a}) shows an exemplary energy deposition simulation, in which the lower noise is visible as less fluctuations between voxel colourisation and a smoother outline out of field resulting from the crop of visualisation at 5\% of the maximum energy deposition. The histograms in Figure~(\ref{fig:stat_unc:b}) show that the statistical uncertainty in the peak areas is below 0.5\% for 97.6\% of the voxels of the low-noise samples (mean value = 0.36\%) while they are less than 2\%  in the valley for 98.1\% of the voxels (mean value = 1.23\%).

\subsection{Machine learning model\label{ML}}
\begin{figure}[t]
	\hspace*{-20mm}
	\includegraphics[width=1.2\textwidth]{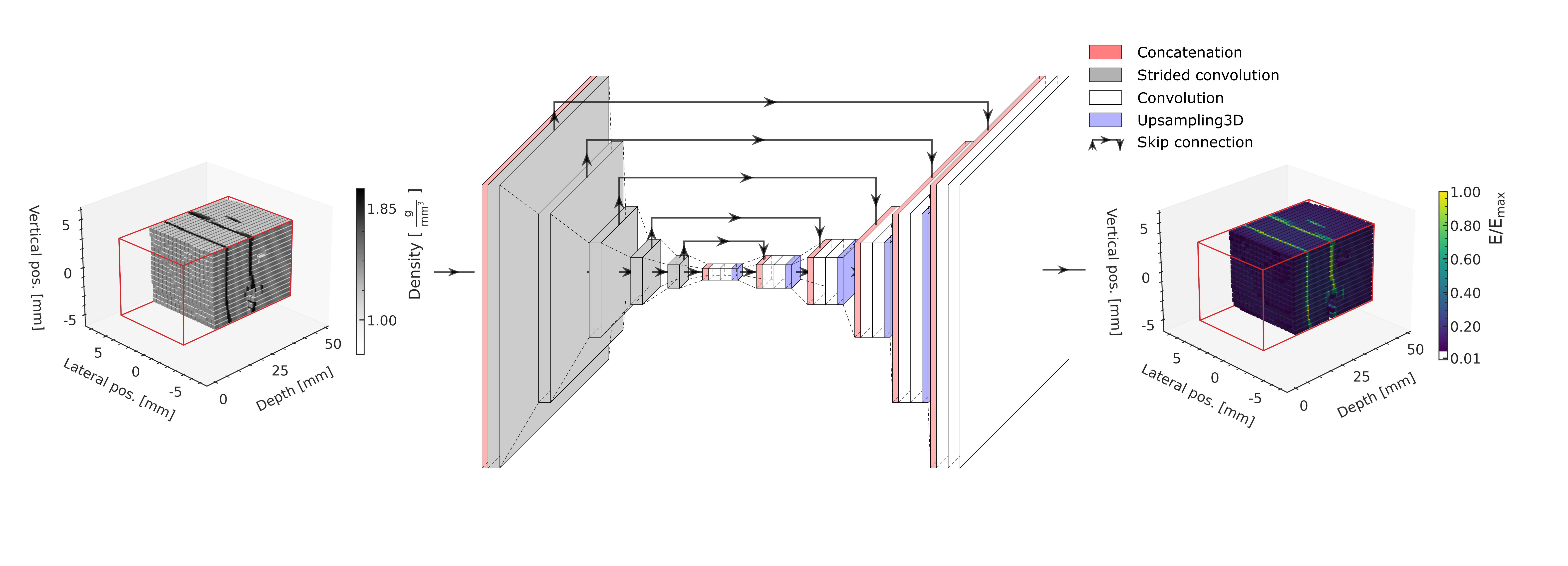}
	\caption{Schematic of the implemented deep learning model predicting energy deposition based on a material matrix input. Adapted from \cite{Mentzel2022}.}
	\label{fig_generator_network}
\end{figure}

\noindent The ML model is the same as in our previously published study~\cite{Mentzel2022} and is based on a 3D U-Net~\cite{Cicek2016}, illustrated in Figure~(\ref{fig_generator_network}). The models are implemented using Tensorflow v2.2~\cite{tensorflow2015-whitepaper}.
The input of the model comprises the 96x16x16 density matrix of the phantom within the prediction volume, indicated with red lines in the schematic. Both the material density (input) and energy deposition (output) matrices are normalized to the range $[0, 1]$ using the respective minimum and maximum values in the training dataset. 
\noindent A \textit{compression path} using strided convolution layers with a following \textit{decompression path} using blocks of 3D upsampling followed by two convolution layers achieves a multi-scale extraction of relevant geometric features from the input to subsequently predict the energy deposition in each voxel. Skip connections between the compression and the decompression path allow bypassing deeper model layers, allowing features of each compression level to be used for the prediction and to avoid vanishing gradients~\cite{He2016}.

\noindent Two independent ML models are trained for the peak and valley energy deposition predictions, respectively. For each of them, an individual search for an optimal \textit{hyperparameter} configuration is conducted by evaluating the performance on the validation data. 

\noindent In the scope of this study, the number of convolution filters per convolution layer, the batch size and the learning rate are varied in the optimisation. For this, different neural networks with the respective settings are trained. For the training, the Adam optimizer~\cite{Kingma2015} is used, together with the mean absolute error (MAE) between the predicted and MC-simulated energy deposition as loss function. The MAE is calculated as $MAE=\frac{\Sigma_{i}^{n} {\lvert y_i -x_i \rvert }}{n}$, where $y_i$ and $x_i$ are the energy depositions calculated by the ML dose engine and the MC simulation in voxel $i$ of the target region with a total number $n$ of voxels.  Training is stopped when the MAE computed on the validation data does not improve anymore for at least 30 epochs. The models of the respective epoch which achieves the lowest MAE on the validation data are used for obtaining the predictions for the test cases and for the corresponding comparisons. 

% All trainings are performed using an Nvidia GForce 1080 Ti graphics processing unit with 11$\,$GB memory. Training durations were between 4 hours and approximately two days, depending on the hyperparameter configuration. 

\subsection{Performance measures}
\noindent In the search for optimal hyperparameters, the MAE computed on the validation data is used as the main measure of comparison. 

\noindent Due to the shifts and rotations used for data augmentation, several data samples exhibit clinically less relevant features like large proportions of the spine or auditory canal, which are both not subject to MRT treatments under current preclinical protocols at the Australian Synchrotron. Especially voxels with bone material lead to large MAE values as the energy depositions are larger within bone structures compared to the brain. To allow for a more outlier-robust comparison of the ML model performance on the training, validation and test datasets, not only the average MAE but also the resulting boxplots are analysed which contain more information about the distribution of deviations.

\noindent Much of the deviation of the ML predictions from high-noise MC data results from statistical fluctuations of the MC simulations themselves. An accurate voxel-wise prediction of energy deposition of high-noise MC simulation data is not only not desired but would also mean poor generalization. Instead, an estimate of the mean value of the underlying energy deposition distribution is desired, which would match an MC simulation of the same scenario with less statistical uncertainty. In order to investigate if the ML model is capable to interpolate the high-noise data, the smooth ML energy deposition predictions are compared to low-noise MC simulation data as well as with the high-noise MC simulation data, relative to their standard error. 

\noindent In the case of an unbiased prediction of the values for each voxel, it is expected that 68\% of values lie within one standard deviation ($1~\sigma$ from the simulation mean value). This expectation value can be used to assess the ML prediction quality in the presence of noisy MC data: if less than 68\% of voxel-wise ML predictions lie within $1~\sigma$ from the MC simulation, the deviations cannot be explained solely by statistical fluctuations, hinting at a systematic deviation of the ML prediction from the simulation. If, on the other side, more than 68\% of the voxel-wise energy depositions agree within $1~\sigma$ between ML and MC models, it points towards an overfitting of the model to the noise present in individual data samples.

\noindent In the case of the three test patient cases with low-noise MC simulations used for the testing, the relative deviations $\Delta D_{rel}=\frac{D_{ML}-D_{MC}}{D_{ML}}$ 
between the ML prediction and MC simulations are assessed, where $D_{ML}$ and $D_{MC}$ are the doses predicted with ML and MC model respectively, in each voxel of the target. 2D visualisations of $\Delta D_{rel}$ are mostly shown in discrete steps in the plots. This is done to allow for an easier visual inspection of results by the reader which is more difficult using continuous colour scales. The agreement between MC and ML predictions is deemed satisfactory when $\Delta D_{rel} < \pm$3\%. This criterion is chosen in agreement with the commonly used 3\% gamma index \cite{Low1998}. However, in contrast to the gamma index, no spatial deviation is allowed as two computational data samples are compared voxel by voxel. 

\noindent In addition, the prediction of the biologically important peak-to-valley dose ratio (PVDR, \cite{Smyth2019}) is compared between ML model and MC simulation.  

\section{Results\label{results}}
\subsection{Hyperparameter optimisation}

\begin{figure}[t]
	\centering
	\begin{subfigure}[t]{0.75\textwidth}
		\includegraphics[width=\linewidth]{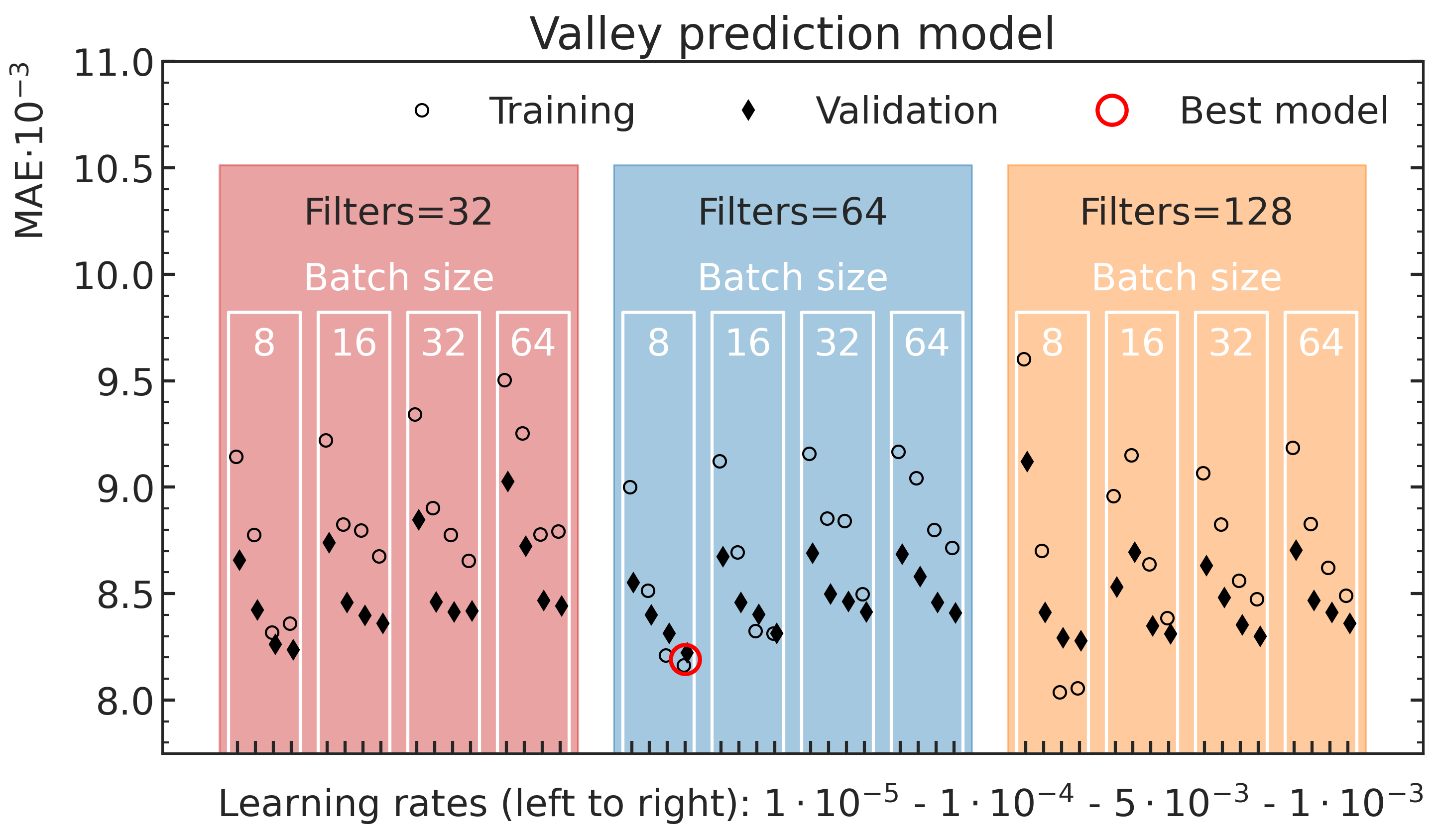}
		\caption{\vspace{.7cm}}
		\label{fig:results:hyperparameter_overview:a}
	\end{subfigure}
	\begin{subfigure}[t]{0.75\textwidth}
		\includegraphics[width=\linewidth]{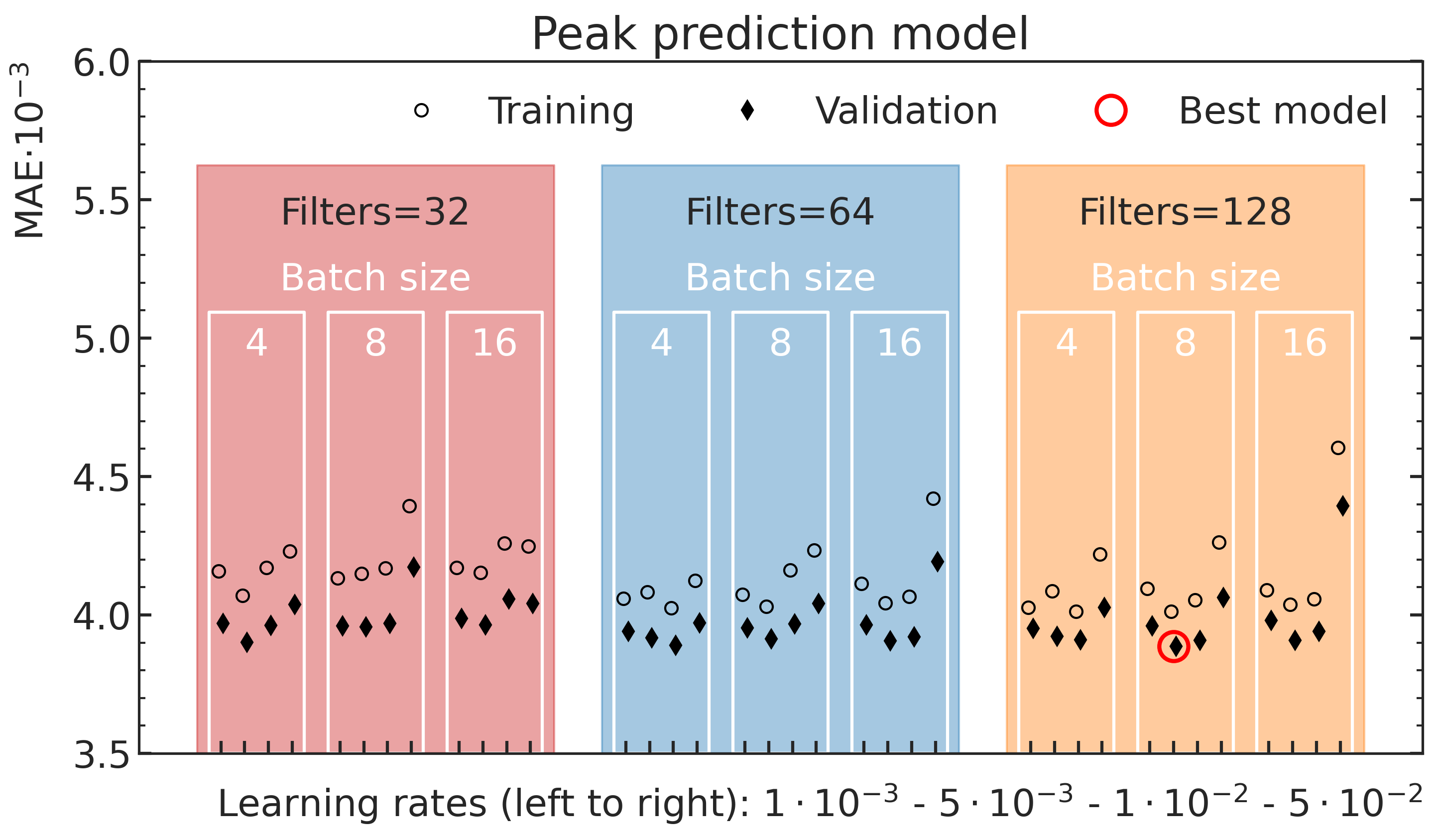}
		\caption{}   
		\label{fig:results:hyperparameter_overview:b}
	\end{subfigure}
	\caption{Overview of the validation loss values (diamonds) and the corresponding training data loss values (open circles) for different valley (a) and peak (b) energy deposition prediction model configurations. The x-ticks locate the different investigated learning rates while the batch sizes and number of filters are highlighted for each model by their positioning in the respective white (batch size) and coloured (number of filters) boxes. The best respective model is marked with a red circle. }
\end{figure}

\noindent The best average MAE on the validation data in dependence on the different hyperparameter settings, together with the corresponding MAE on the training data, are shown for the valley model in Figure~(\ref{fig:results:hyperparameter_overview:a}) and the peak model in Figure~(\ref{fig:results:hyperparameter_overview:b}). In the prediction of the valley energy deposition, the model with 64 convolution filters in each convolutional layer, a batch size of 8 and a learning rate of  $1 \times 10^{-3}$ resulted in the best validation performance. Training with smaller batch sizes or larger learning rates did not converge. The best model for the peak predictions used 128 convolutional filters in each convolution layer, a batch size of 8 and a learning rate of  $5 \times 10^{-3}$. Although training with smaller batch sizes and larger learning rates did converge for these training runs, no better results were achieved. 

%\noindent 
%\begin{figure}[t]
	%	\centering
	%	\hfill
	%	\begin{subfigure}[t]{0.45\textwidth}
		%		\includegraphics[width=\linewidth]{figure_11_1.png}
		%		\caption{}
		%		\label{fig:results:loss_curves:a}
		%	\end{subfigure}
	%	\hfill
	%	\begin{subfigure}[t]{0.45\textwidth}
		%		\includegraphics[width=\linewidth]{figure_9_1.png}
		%		\caption{}
		%		\label{fig:results:loss_curves:b}
		%	\end{subfigure}
	%	\hfill
	%	\caption{Loss curve of the best valley (a) and peak (b) prediction model indicating the epoch in which the final model is chosen for the peak prediction.}
	%	\label{fig:results:loss_curves}
	%\end{figure}
%
%\noindent The MAE loss curves including the epoch resulting in the best validation loss for the valley and peak energy deposition models are shown in Figure~(\ref{fig:results:loss_curves:a}) and  Figure~(\ref{fig:results:loss_curves:b}), respectively. In the case of the valley model, the training loss continues to decrease while the validation loss stagnates, hinting at potential overtraining for longer than the chosen number of epochs but does not show indications for overtraining in the chosen epoch. For the peak model, the training loss is generally larger than the validation loss. This effect trend that can also be observed in many hyperparemter-dependent results in Figure~(\ref{fig:results:hyperparameter_overview}). 

\subsection{Performance and generalisation assessment\label{results:generalization}}

\begin{figure}[t]
	\hspace*{-20mm}
	\centerline{
		\begin{subfigure}[t]{0.43\textwidth}
			\includegraphics[width=\linewidth]{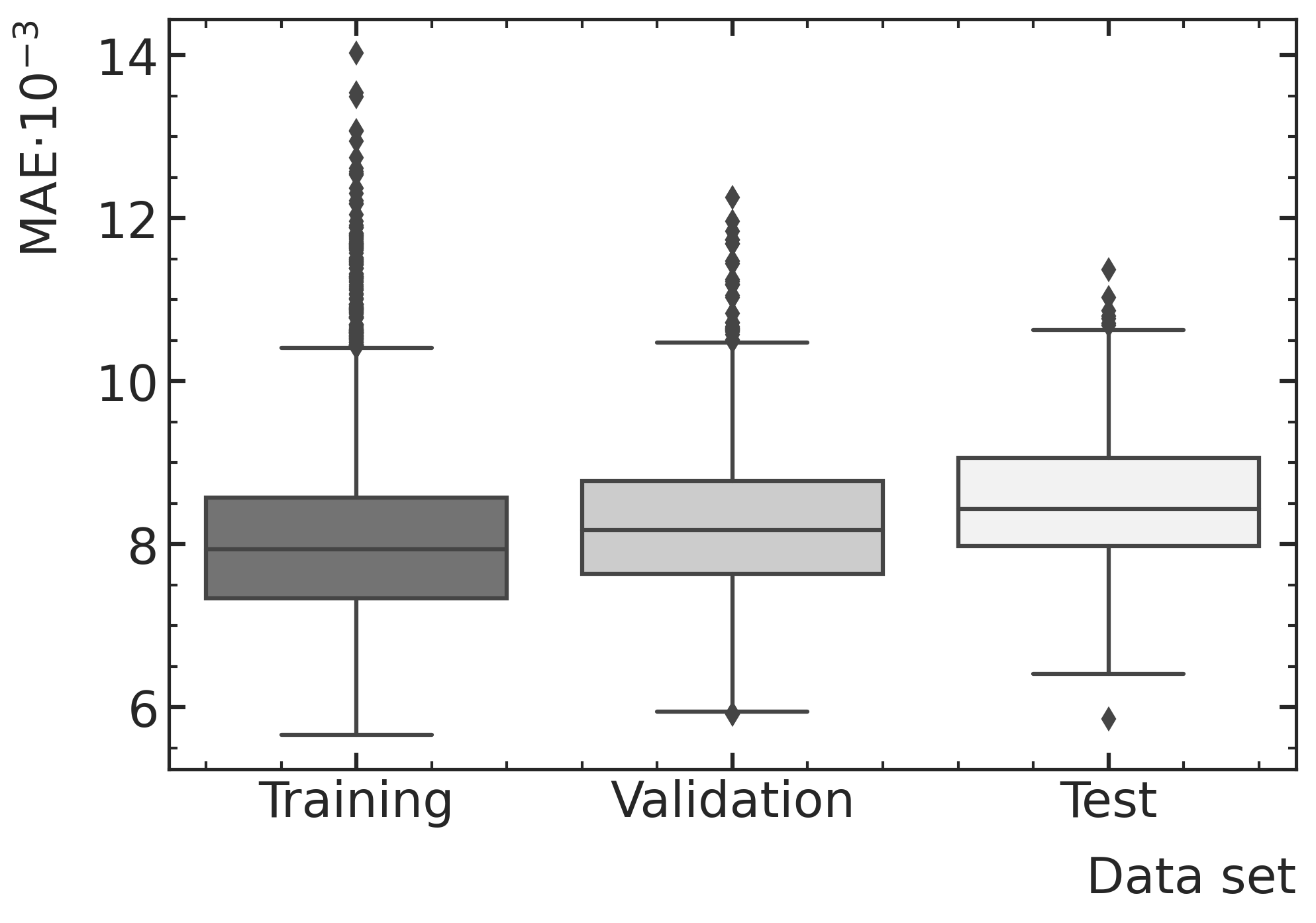}
			\caption{}
			\label{fig:results:valley_performance:a}
		\end{subfigure}
		\begin{subfigure}[t]{0.4\textwidth}
			\includegraphics[width=\linewidth]{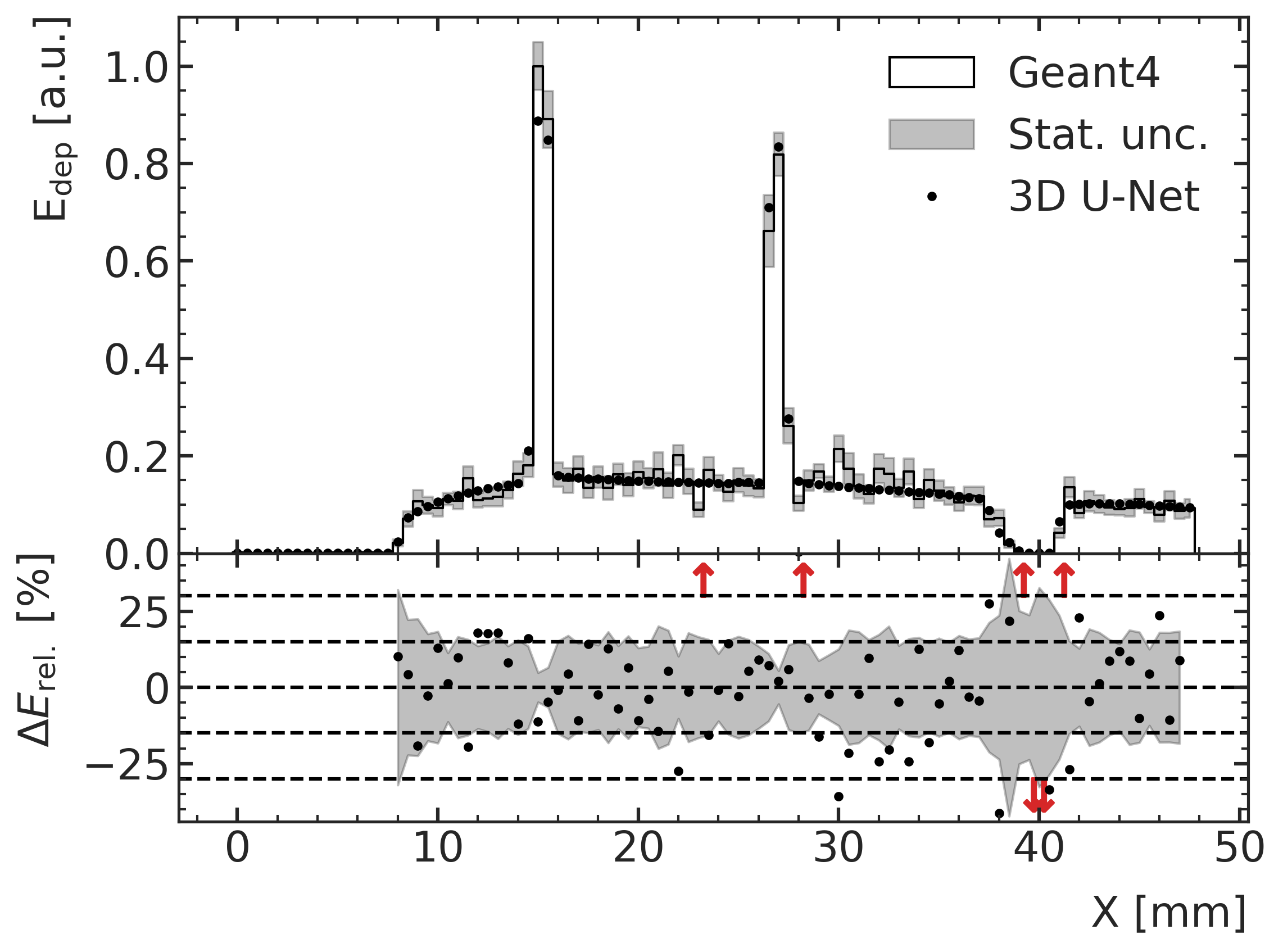}
			\caption{}
			\label{fig:results:valley_performance:b}
		\end{subfigure}
		\begin{subfigure}[t]{0.43\textwidth}
			\includegraphics[width=\linewidth]{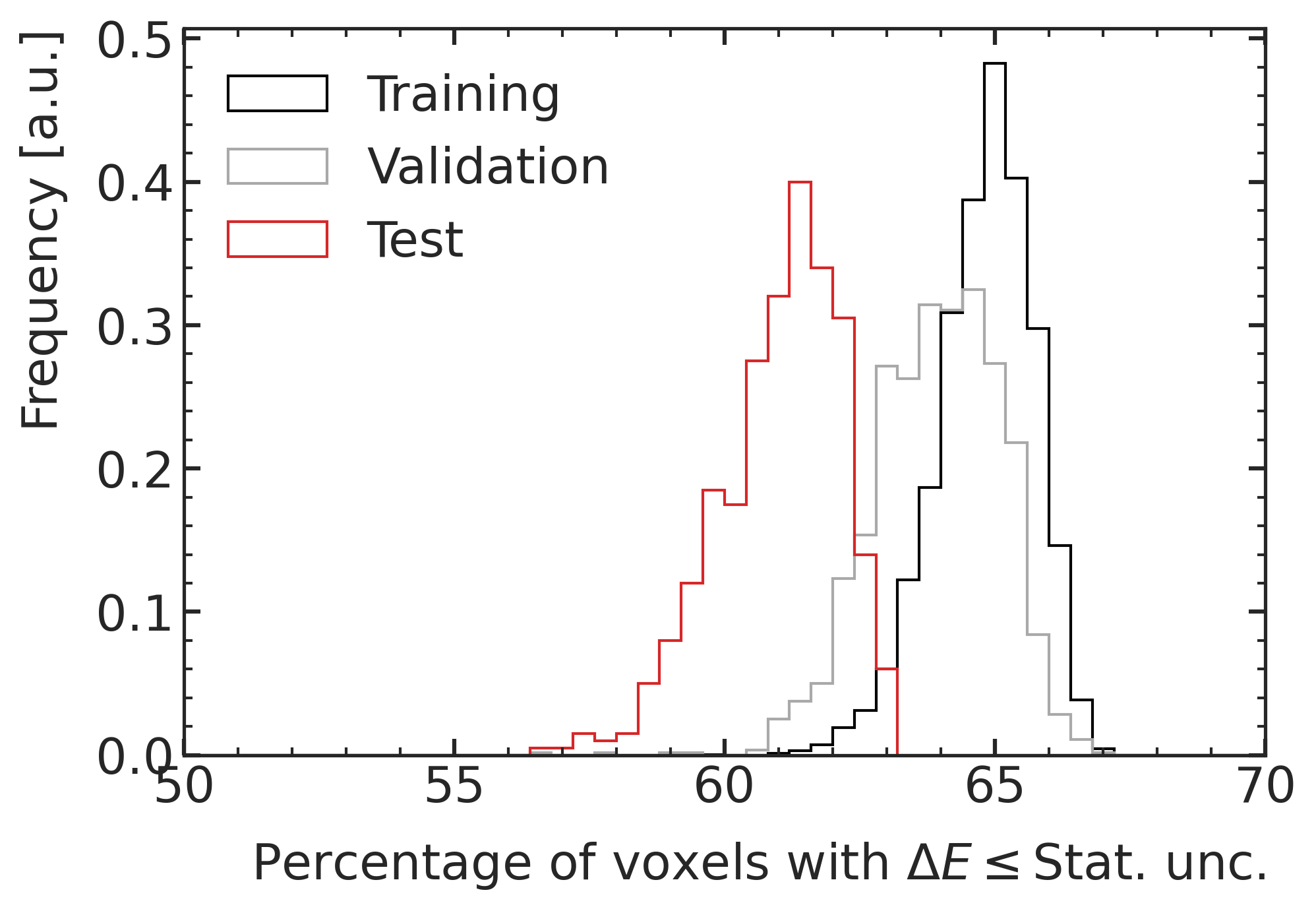}
			\caption{}
			\label{fig:results:valley_performance:c}
	\end{subfigure}}
	\caption{(a) Boxplots showing the MAE in the valley region, for the training, validation and test datasets. The central line of the each boxplot shows the median of the distribution. The surrounding box is limited by the 25\% percentile. The whiskers are shown at 2.5$\times$25\% percentile. Data samples further away from the median are represented as outliers. (b) Exemplary ML predicted and  MC simulated energy deposition $E_\mathrm{dep}$ of the validation data in the valley region. The bottom plot shows the percent relative difference $\Delta E_\mathrm{rel}$ between ML prediction and MC simulation in terms of energy deposition. Red arrows in the relative energy deviation subplot indicate deviations larger than the shown ranges. (c) Fraction of voxels of the ML predicted energy deposition maps exhibiting a deviation of one standard deviation or less with respect to the mean energy deposition $\Delta E$ calculated with the MC simulation.}
	\label{fig:results:valley_performance}
\end{figure}

\begin{figure}[t]
	\hspace*{-20mm}
	\centerline{
		\begin{subfigure}[t]{0.43\textwidth}
			\includegraphics[width=\linewidth]{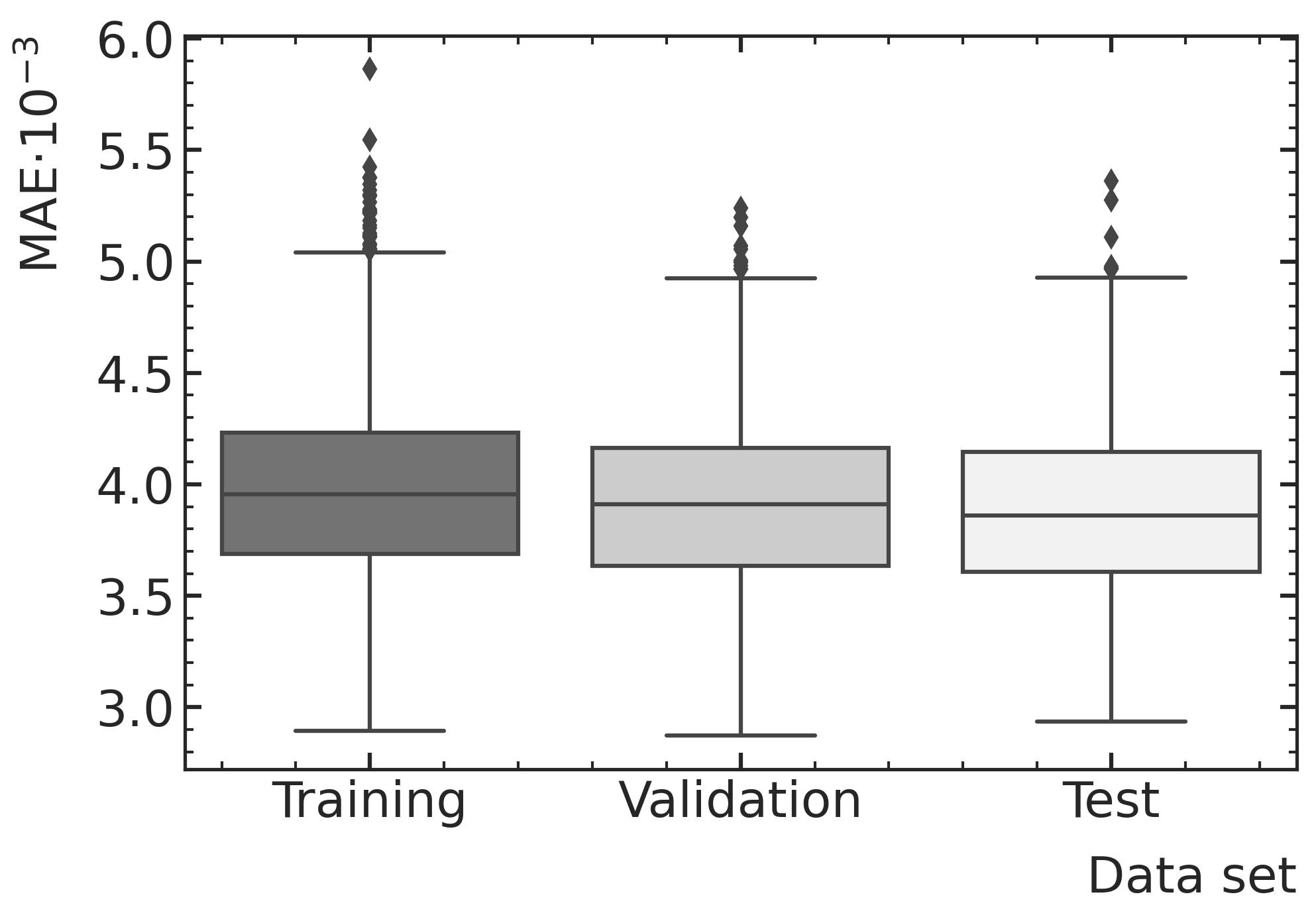}
			\caption{}
			\label{fig:results:peak_performance:a}
		\end{subfigure}
		\begin{subfigure}[t]{0.4\textwidth}
			\includegraphics[width=\linewidth]{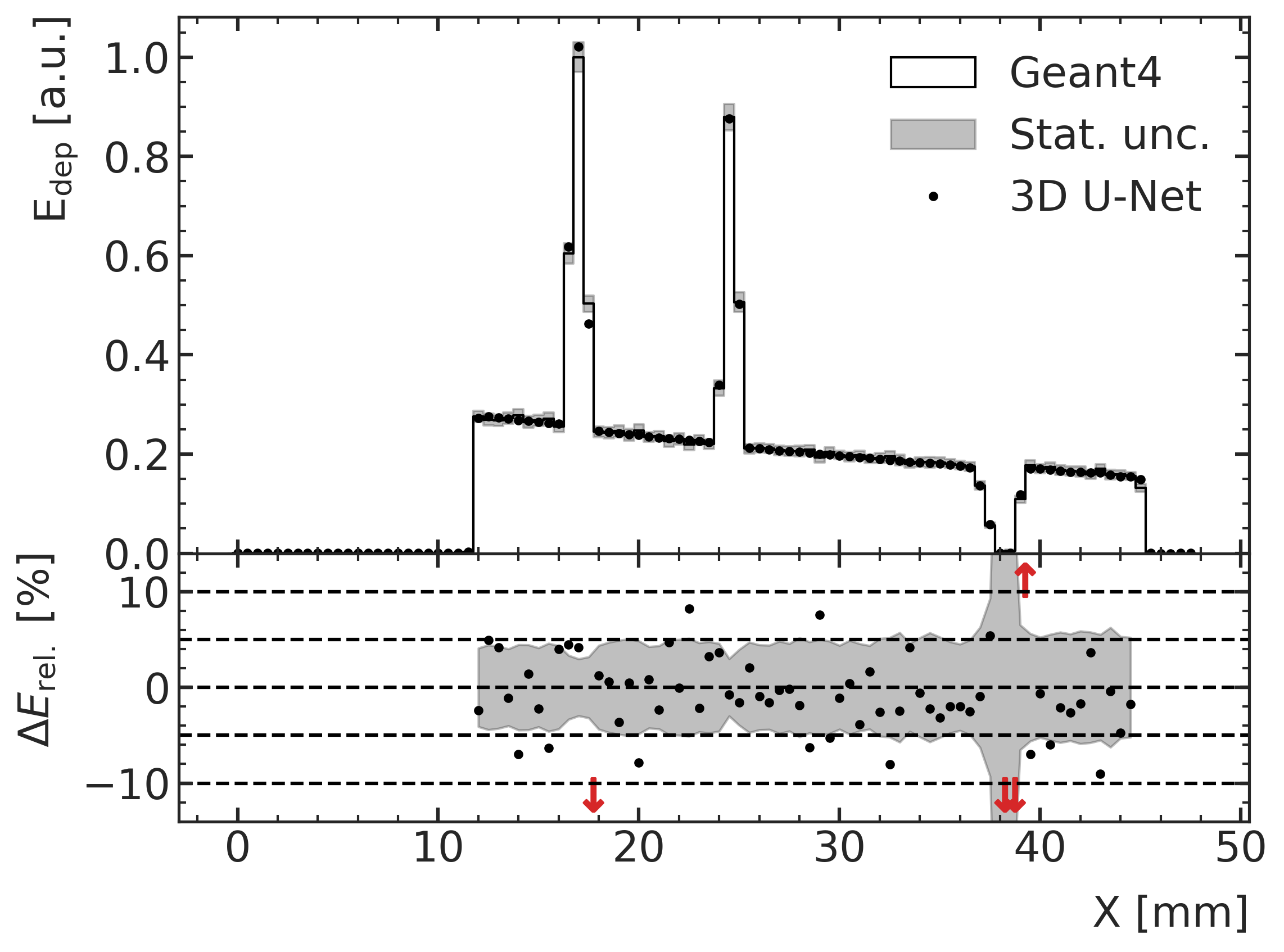}
			\caption{}
			\label{fig:results:peak_performance:b}
		\end{subfigure}
		\begin{subfigure}[t]{0.43\textwidth}
			\includegraphics[width=\linewidth]{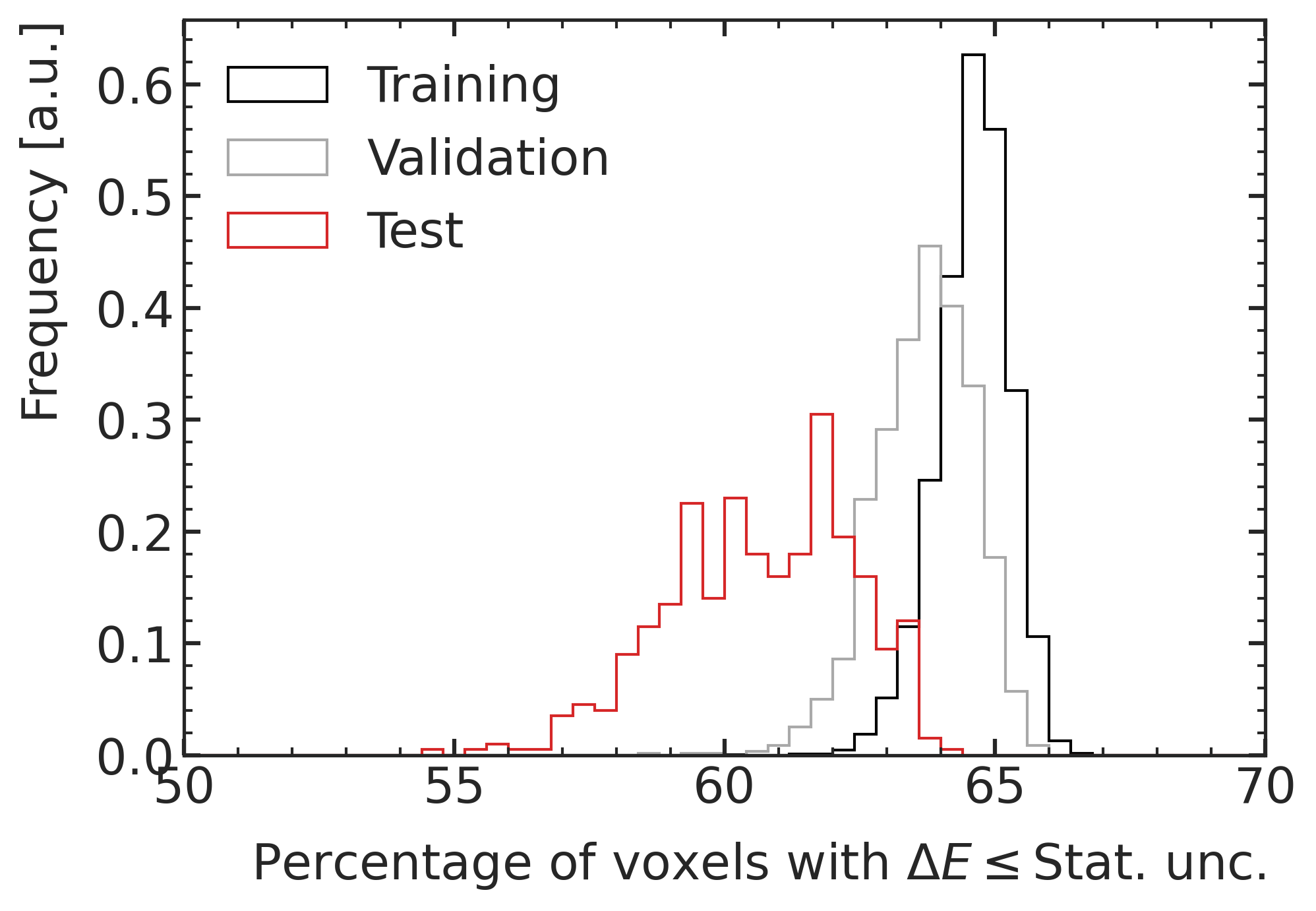}
			\caption{}
			\label{fig:results:peak_performance:c}
	\end{subfigure}}
	\caption{(a) Boxplots showing the MAE in the peak region for the training, validation and test datasets. The central line of the each boxplot shows the median of the distribution. The surrounding box is limited by the 25\% percentile. The whiskers are shown at 2.5$\times$25\% percentile. Data samples further away from the median are represented as outliers. (b) Exemplary ML predicted and MC simulated energy deposition $E_\mathrm{dep}$ of the validation data in the peak region. The bottom plot shows the percent relative difference $\Delta E_\mathrm{rel}$ between ML prediction and MC simulation in terms of energy deposition. Red arrows in the relative energy deviation subplot indicate deviations larger than the shown ranges.(c) Fraction of voxels of the ML predicted energy deposition maps exhibiting a deviation of one standard deviation or less with respect to the mean energy deposition $\Delta E$ calculated with the MC simulation.}
	\label{fig:results:peak_performance}
\end{figure}

The optimal peak and valley ML models are used to predict all high-noise training, validation and test data samples to assess the overall performance and generalisation.
Figure~(\ref{fig:results:valley_performance}) and~(\ref{fig:results:peak_performance}) show examples of results for the valley and peak regions, respectively. 
Figure~(\ref{fig:results:valley_performance:a}) and~(\ref{fig:results:peak_performance:a}) show 
boxplots of the MAE for the training, validation and test data. Figure~(\ref{fig:results:valley_performance:b}) and~(\ref{fig:results:peak_performance:b}) illustrate depth-energy deposition curves at the centre of the microbeam field for one exemplary validation data sample for both MC and ML models. The predictions of the ML model agree well with the simulated data within the statistical uncertainty of the MC simulation, while being significantly smoother, which contributes to the assessment that the model can generalise to unseen test data. Larger deviations can be seen in areas of very low density as occurring in both samples deeper in the phantom. 
\noindent A closer investigation into the generalisation and performance is done by analysing the fraction of voxels, for which the deviation between the ML and MC prediction is smaller than 1~$\sigma$ of the statistical uncertainty of the simulation. As shown in Figure ~(\ref{fig:results:valley_performance:c}) and (\ref{fig:results:peak_performance:c}), the distribution of the training data peaks at a value around 65\% closest to the expected value of 68\%, which means the deviations are mostly of pure statistical nature. While the distribution of the validation data is only slightly broader and shifted to lower values, the distribution of the test data averages around 61\% and is visibly broader for the peak predictions, which indicates that the model does not generalize perfectly to the unknown geometries of the test data. 

\noindent The averaged fractions of voxels with deviations smaller than 1~$\sigma$ are shown together with the averaged MAE for all three datasets in Table (\ref{tab:mae_results}). While the averaged MAE agrees well within uncertainties between the three datasets, the averaged fraction of voxels indicate a small bias on a voxel-by-voxel basis if the model is evaluated with independent data.

\begin{table}[t]
	\centering
	\renewcommand{\arraystretch}{1.2}
	\begin{center}
		\caption{Average MAE and fraction of voxels with an absolute dose difference ($\Delta$D) between MC and ML calculation of less than $1~\sigma$, computed for the peak and valley predictions. The datasets derive from the MC training, validation and testing. The mean MAEs and associated standard errors are calculated from the MAEs obtained for the augmented data of the 16 rats (10 for training, 3 for validation and 3 for testing). For the voxel fractions (second and forth columns), the reported mean values and standard deviations are calculated considering all the datasets used in the various cases under study, determining the mean value and standard deviation from  the individual distributions of each dataset represented in Fig.~(\ref{fig:results:valley_performance:c}) and~(\ref{fig:results:peak_performance:c}) for an exemplary case.} \label{tab:mae_results}
		\begin{tabularx}{\textwidth}{>{\raggedright\arraybackslash}X >{\centering\arraybackslash}X >{\centering\arraybackslash}X >{\centering\arraybackslash}X >{\centering\arraybackslash}X }	
			\toprule
			& Valley & & Peak & \\
			\midrule
			Dataset & MAE [$1\times10^{-3}$] & $\Delta $D < Stat. unc.  [\%] & MAE [$1\times10^{-3}$] & $\Delta $D < Stat. unc. [\%]\\
			\midrule
			Training & $8.2\pm0.3$ & $64.8\pm0.9$ & $4.0\pm0.2$ & $64.6\pm0.7$\\
			Validation & $8.2\pm0.2$ & $63.9\pm1.2$ & $3.9\pm0.1$ & $63.7\pm0.9$\\
			Test & $8.4\pm0.1$ & $61.0\pm1.1$ & $4.1\pm0.1$ & $60.7\pm1.7$\\
			\bottomrule
		\end{tabularx}
	\end{center}
\end{table}

\noindent The training loss is observed to be higher than the validation loss at least for the chosen peak prediction model and this tendency is visible for multiple valley prediction models in Figure~(\ref{fig:results:hyperparameter_overview:a}) as well. This is a result of simulation samples from the rats with numbers 1 and 8 (both in the training dataset) exhibiting a larger number of samples than average that include a relevant proportion of spine in the path of the beam. 

\begin{figure}[t]
	\centering
	\hspace*{-20mm}
	\centerline{
		%\hfill
		\begin{subfigure}[t]{0.4\textwidth}
			\includegraphics[width=\linewidth]{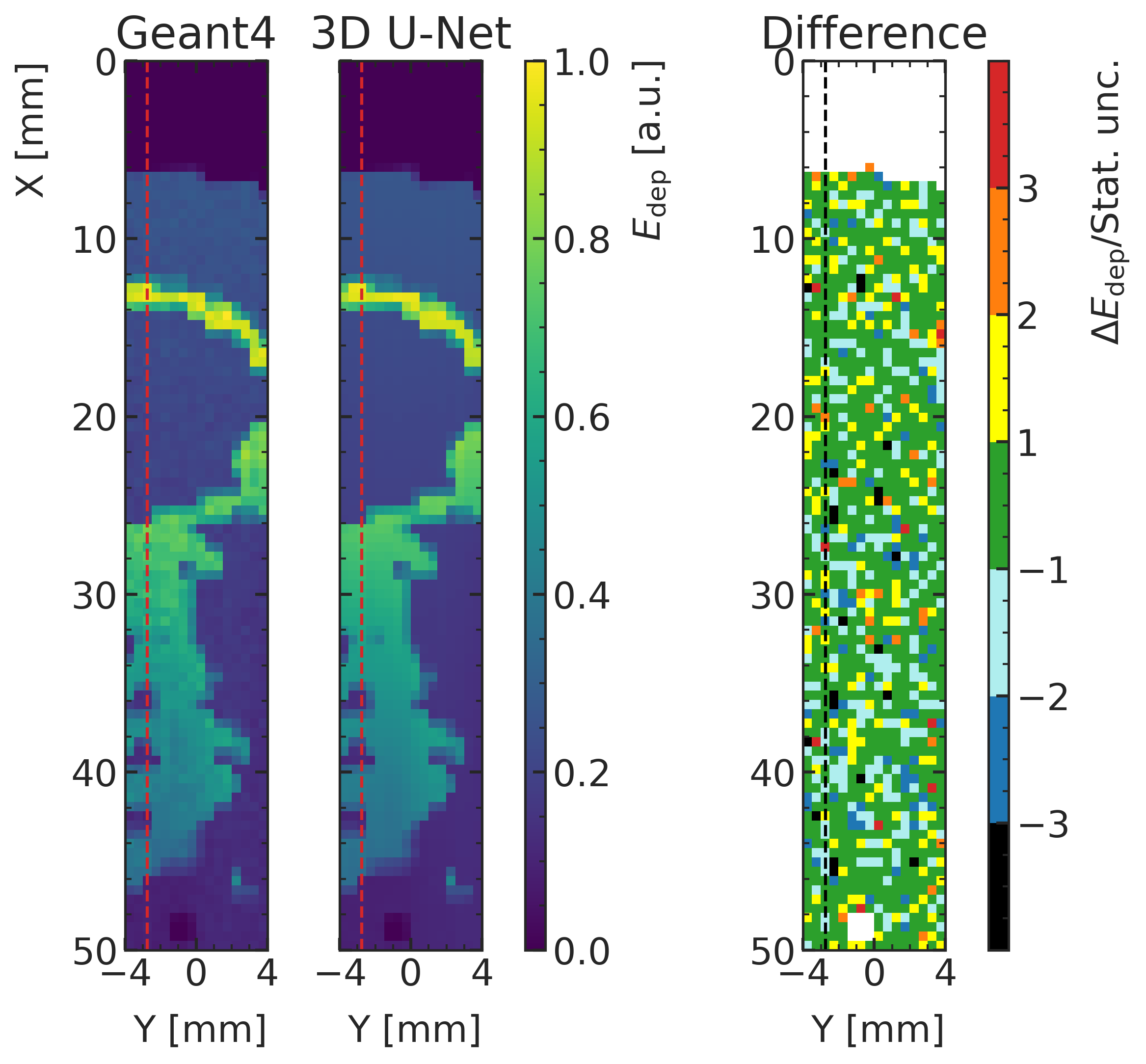}
			\caption{\vspace{.5cm}}
			\label{fig:spine_worst_example_peak:a}
		\end{subfigure}
		\hfill
		\begin{subfigure}[t]{0.4\textwidth}
			\includegraphics[width=\linewidth]{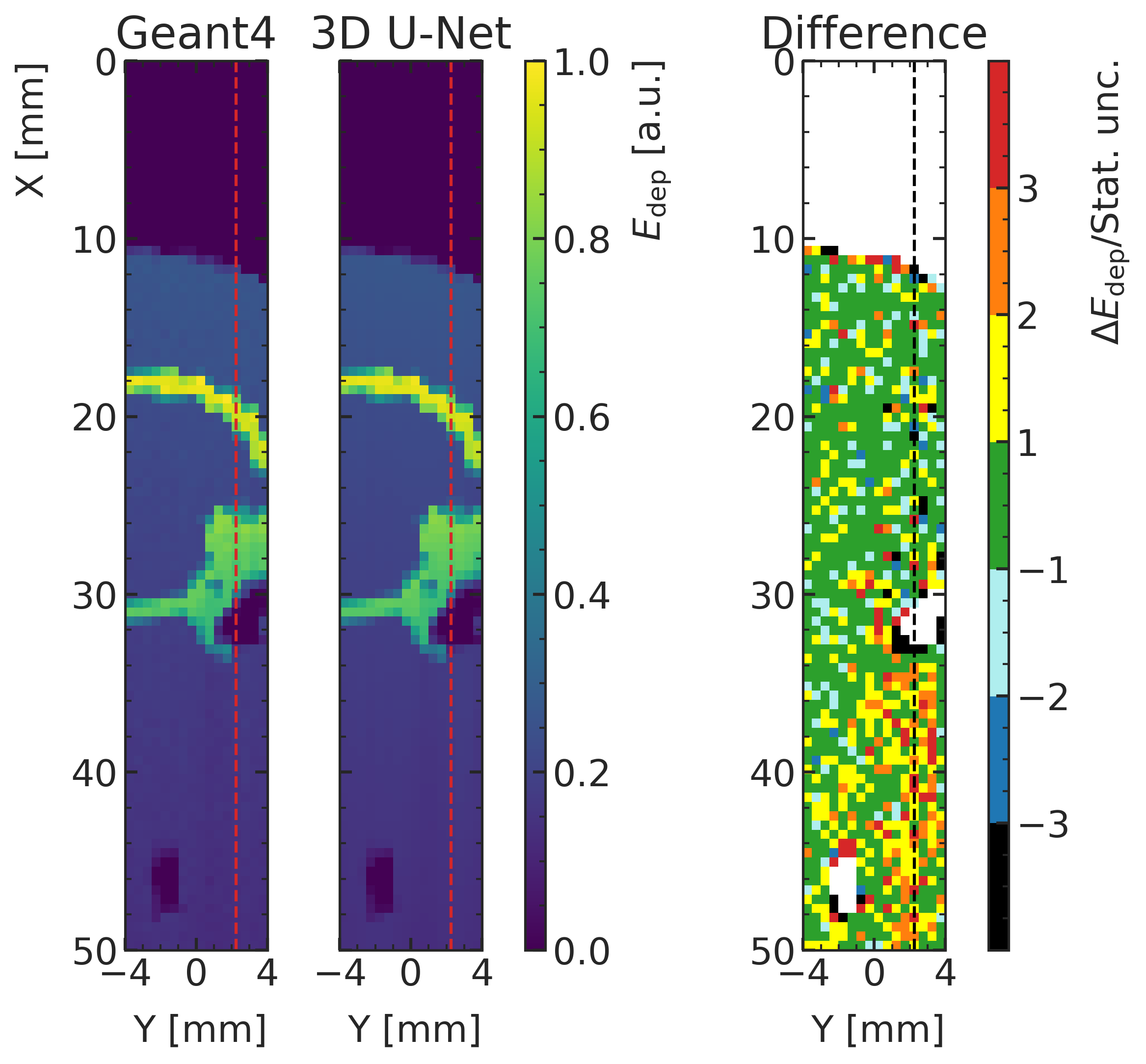}
			\caption{\vspace{.5cm}}
			\label{fig:spine_worst_example_peak:b}
		\end{subfigure}
		\hfill
		\begin{subfigure}[t]{0.4\textwidth}
			\includegraphics[width=\linewidth]{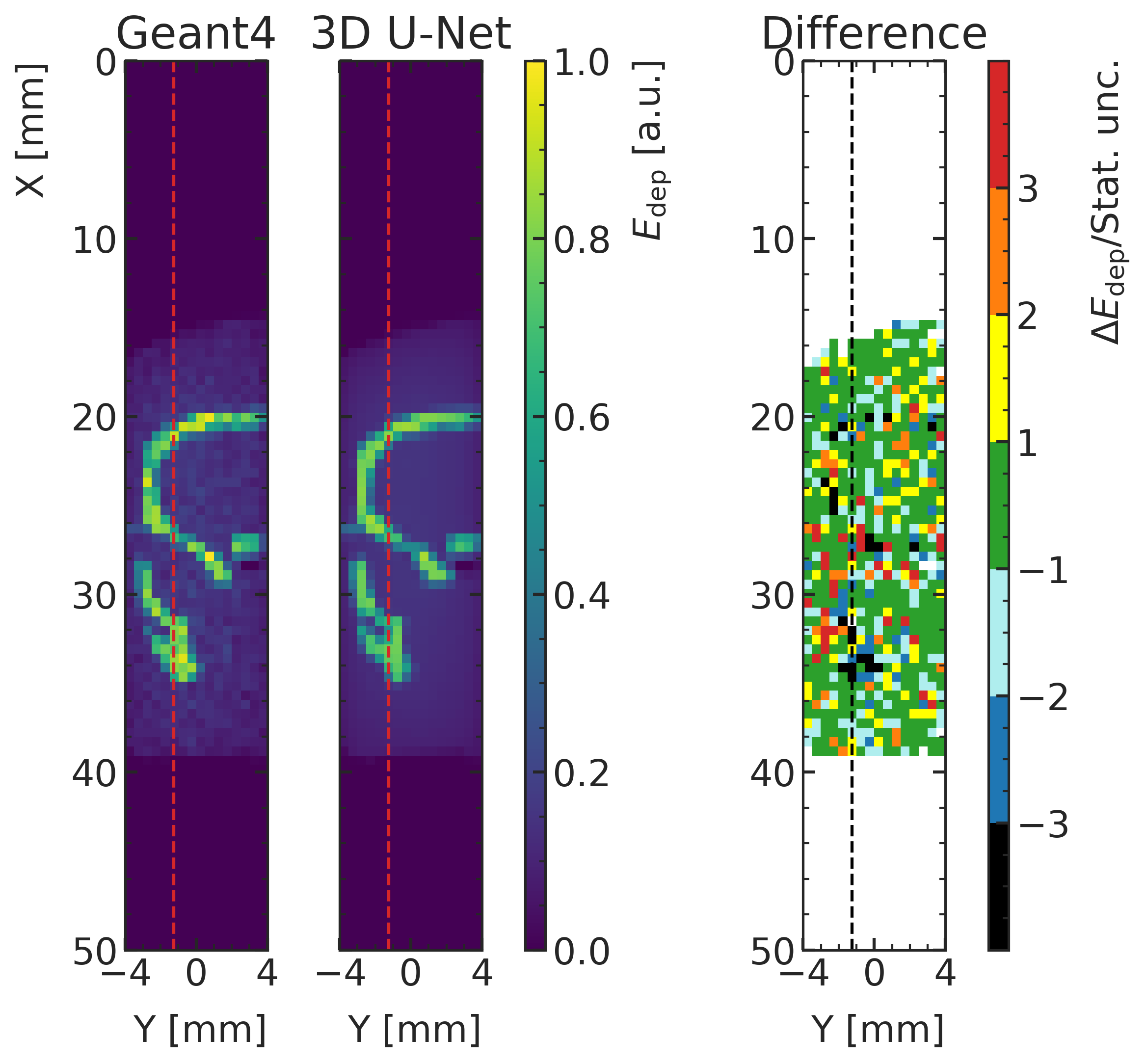}
			\caption{\vspace{.5cm}}
			\label{fig:spine_worst_example_peak:c}
		\end{subfigure}
	}
	\hspace*{-20mm}
	\centerline{
		\begin{subfigure}[t]{0.4\textwidth}
			\includegraphics[width=\linewidth]{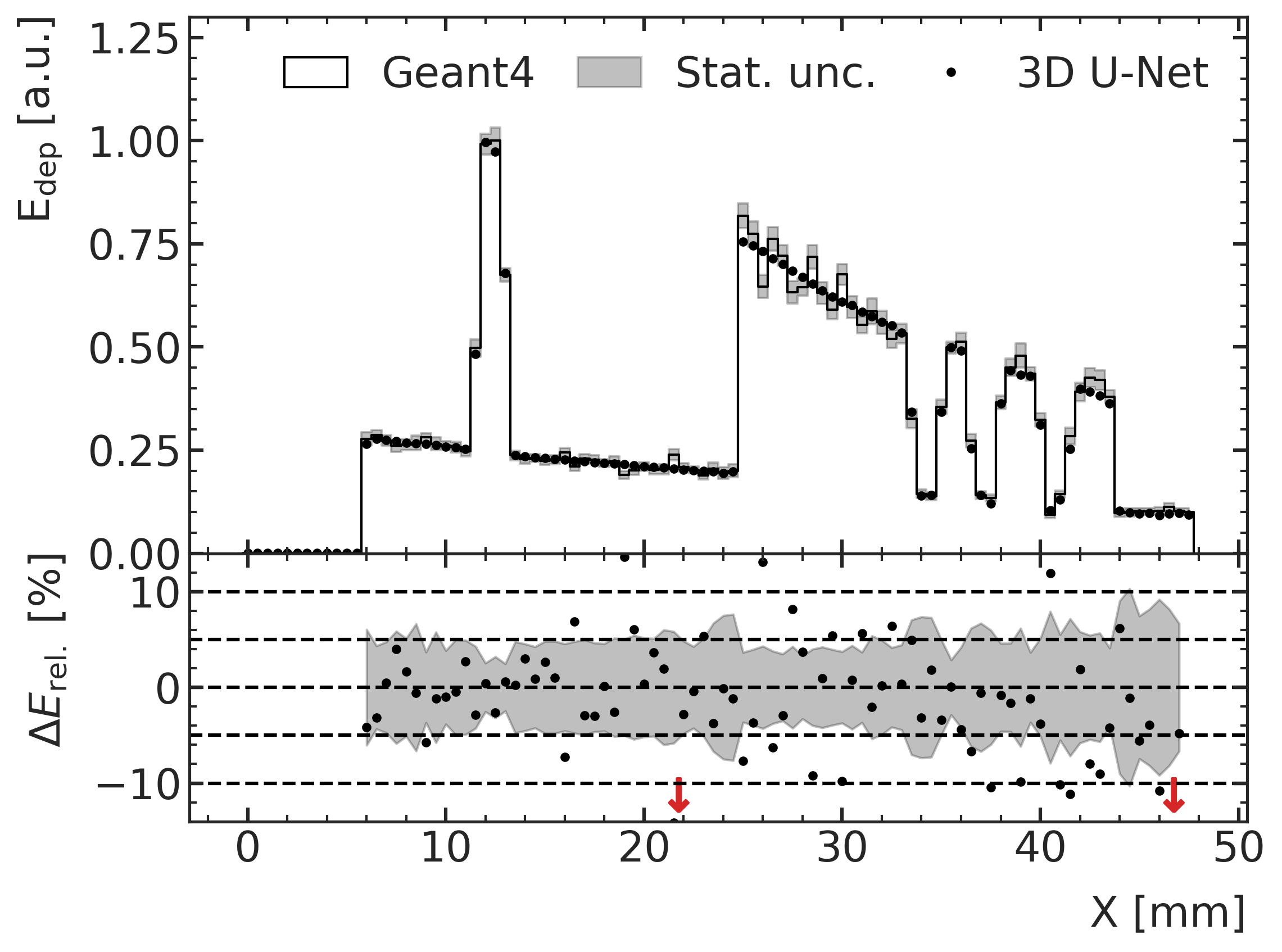}
			\caption{}
			\label{fig:spine_worst_example_peak:d}
		\end{subfigure}
		\hfill
		\begin{subfigure}[t]{0.4\textwidth}
			\includegraphics[width=\linewidth]{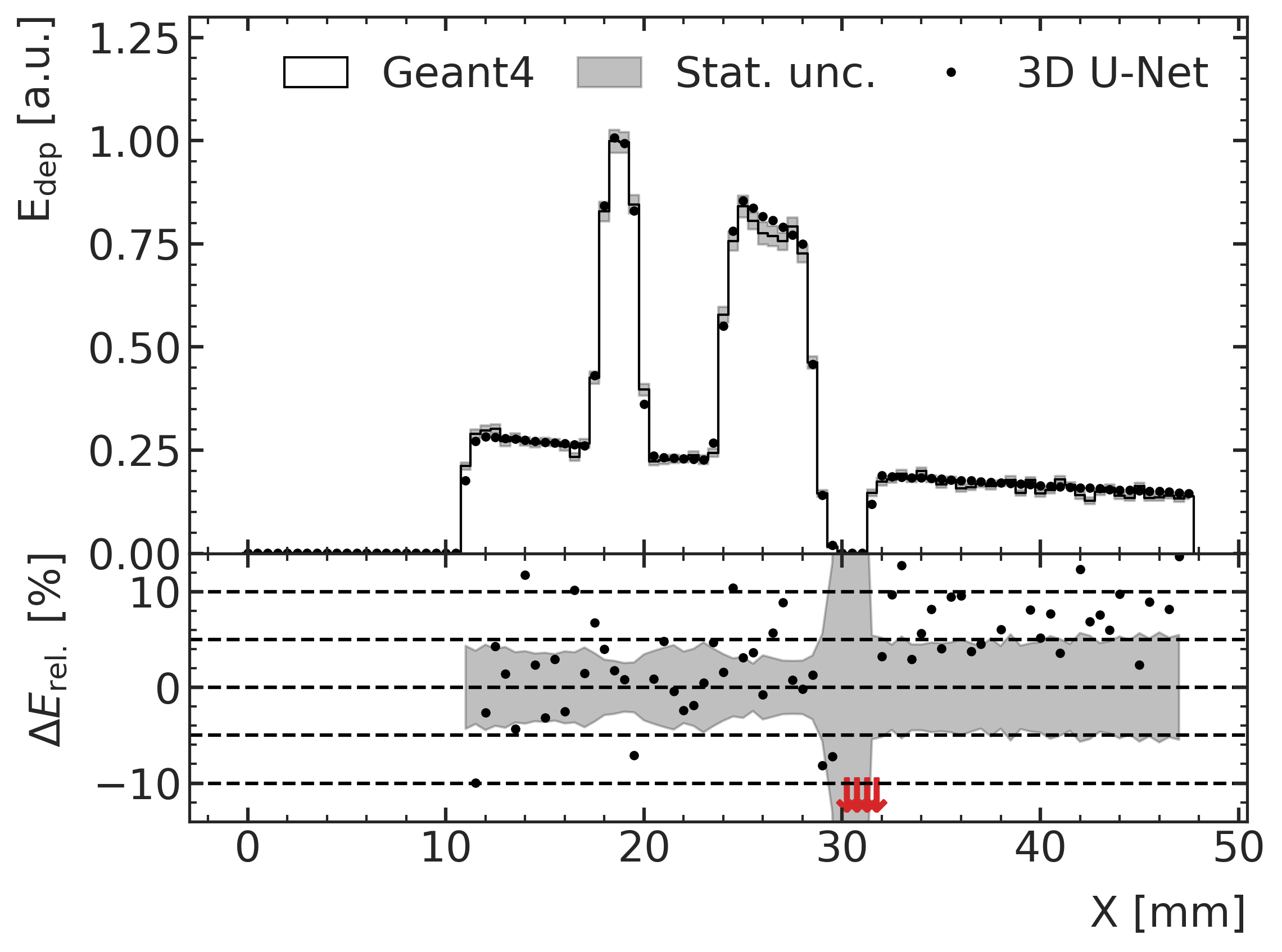}
			\caption{}
			\label{fig:spine_worst_example_peak:e}
		\end{subfigure}
		\hfill
		\begin{subfigure}[t]{0.4\textwidth}
			\includegraphics[width=\linewidth]{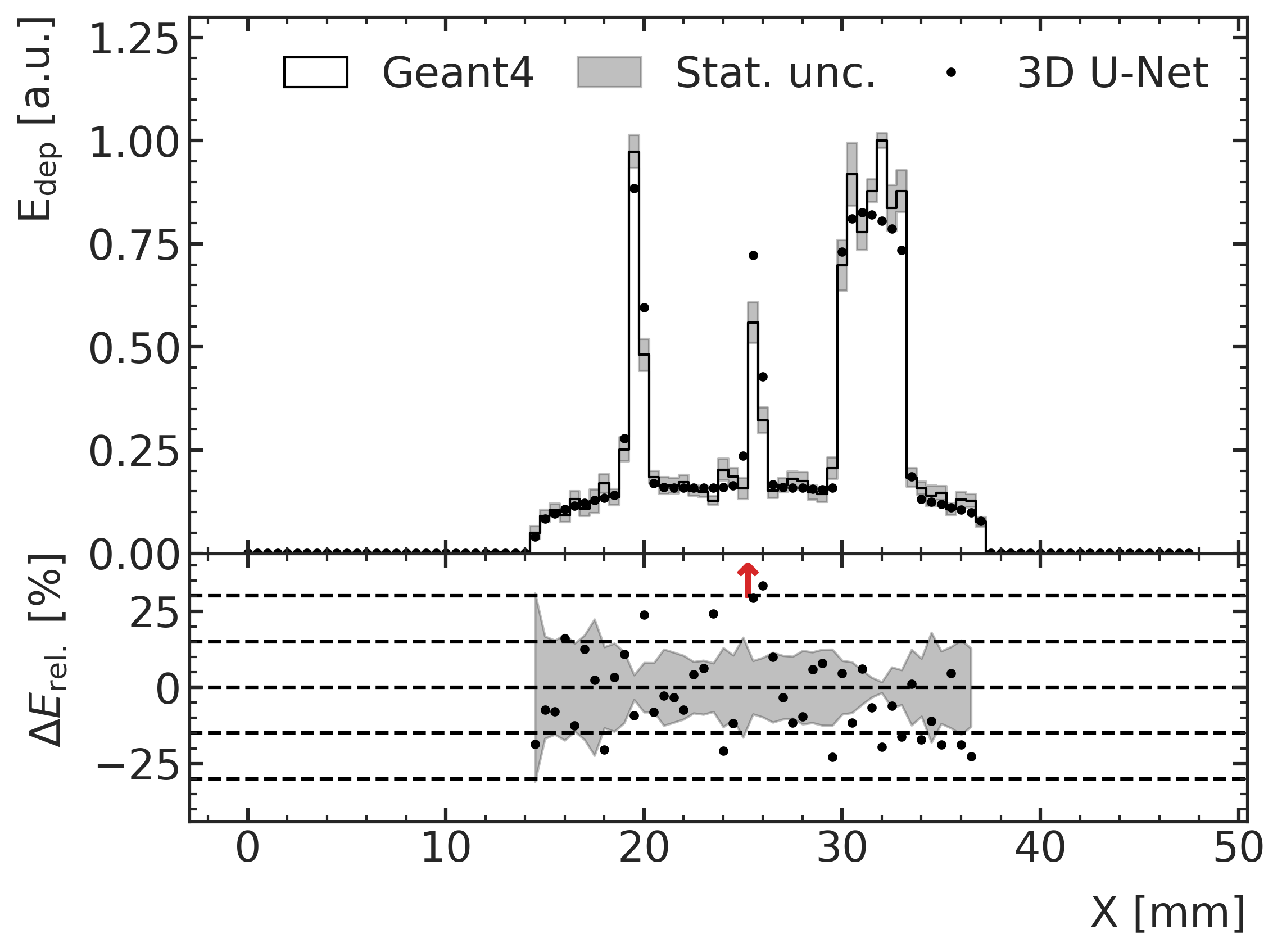}
			\caption{}
			\label{fig:spine_worst_example_peak:f}
		\end{subfigure}
	}
	\hfill
	%	\begin{subfigure}[t]{0.42\textwidth}
		% 		\includegraphics[width=\linewidth]{peak_worst_example_spine_depth_edep}
		%		\caption{}
		%		\label{fig:spine_worst_example_peak:b}
		% 	\end{subfigure}
	\caption{(a) Exemplary peak prediction of a training data sample including a larger proportion
		of spine, showing a 2D slice of MC simulation and ML prediction with the difference in units of
		statistical standard deviations. (b) and (c) show two worst-case prediction cases following different
		criteria. (b) Test data sample with the largest average deviation between ML and MC in units of
		standard deviations, in the peak region. (c) Test data sample with the lowest fraction of voxels in which ML prediction
		with MC simulation agree within one standard deviation, in the valley. (d-f) The depth-energy deposition curve at
		the position indicated with red (black) dashed for each 2D representation shown in (a)-(c). Red arrows
		in the relative energy deviation subplot below indicate deviations larger than the shown ranges.}
	\label{fig:spine_worst_example_peak}
\end{figure}

\noindent In Figure~(\ref{fig:spine_worst_example_peak:a}) an exemplary prediction of a training data sample with a large proportion of bone is shown as a 2D slice and is compared to the MC simulation relative to its statistical uncertainty. The corresponding depth profile, indicated with a red (black) dashed line in the 2D slices of the energy predictions (relative differences), is shown in Figure~(\ref{fig:spine_worst_example_peak:d}). Even though this case is not representative of the treatment field used in this preclinical work, the model is capable to predict the energy depositions quite accurately despite the large gradients in energy.
In the bone voxels, there is more energy deposition, therefore the absolute differences in this physical quantity are larger than those calculated in water, although the relative differences are the same, as shown in Figure~(\ref{fig:spine_worst_example_peak:a}). This results in larger MAE for samples comprising a larger number of bone voxels. However, when comparing the performance on the different datasets using boxplots (see Fig.~(\ref{fig:results:valley_performance:a}) and (\ref{fig:results:peak_performance:a})) instead of the mean MAE value only, which are more robust against outliers, the effect of larger absolute differences in the energy deposition in the bone is less significant or not visible at all.

%\noindent A closer investigation into the generalisation and performance is done by analysing the distribution of voxel-wise energy deposition deviations between ML and MC, shown in Figure~\ref{fig:results:valley_performance:c} and Figure~\ref{fig:results:peak_performance:c}: It can be seen that for both models, the prediction of training data matches most closely the expected average of 68\% of voxels exhibiting a deviation in predicted energy deposition smaller than one standard deviation of statistical MC uncertainty. The averages for both models on all three datasets are shown in Table~(\ref{tab:mae_results}). For both models, on average approximately 65\% of the voxel-wise energy deposition predictions are in agreement with the MC simulation within one standard deviation. While the ratio of predictions in agreement for the validation samples is not significantly different from the training dataset, around 61\% of the voxel-wise ML predictions are in agreement with the MC simulations in case of the test data prediction for both models. 
%

The two examples of the test data with the respective lowest agreement between ML prediction and MC simulation are shown in Figure~(\ref{fig:spine_worst_example_peak:b}) and~(\ref{fig:spine_worst_example_peak:c}) for the peak and valley region, respectively. 
Figure~(\ref{fig:spine_worst_example_peak:d}) and Figure~(\ref{fig:spine_worst_example_peak:e}) show one depth-energy deposition curve for each 
of these samples at a position indicated by red dashed lines in the 2D visualisations. In 
the case of the peak model, a systematic overestimation of the energy deposition behind 
an air pocket in the phantom (auditory channel) can be seen. In the case of the valley 
model, especially predictions of relatively thin bone structures lead to lower agreement to the 
MC data. Despite the fact that these are extreme cases and clinically irrelevant cases for MRT, the model still does a reasonably satisfactory job in predicting the deposited energies. 

\subsection{Predictions for test rat patients\label{results:patients}}

\begin{figure}[t]
	\hspace*{-20mm}
	\centering
	\begin{subfigure}[t]{0.35\textwidth}
		\includegraphics[width=\linewidth]{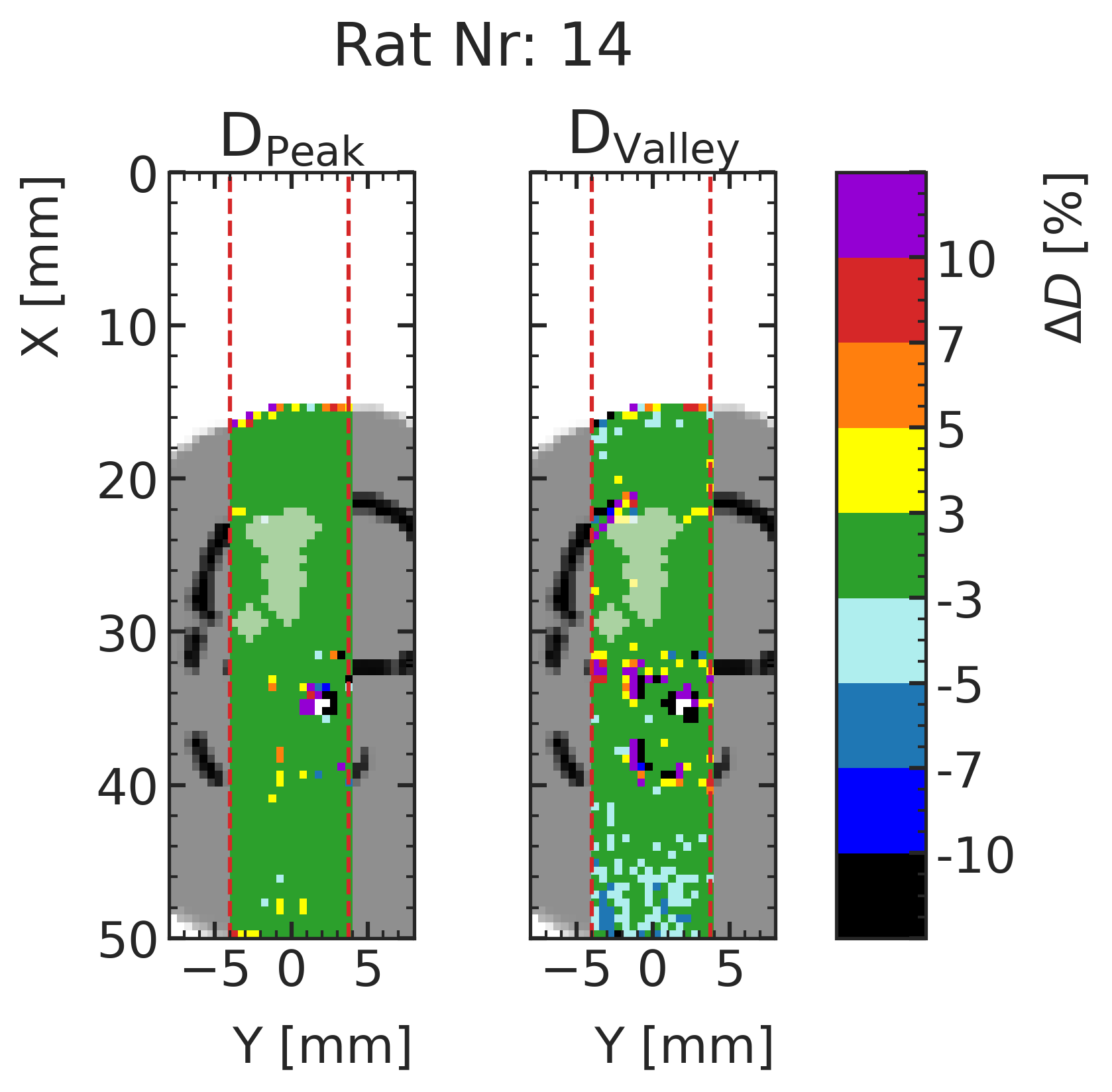}
		\caption{\vspace{.2cm}}
		\label{fig:results:rel_deviations_2d:a}
	\end{subfigure}
	%	\begin{subfigure}[t]{0.32\textwidth}
		%		\includegraphics[width=\linewidth]{GRAY_WITHTUMOR_exemplary_data_2d_rat15_with_DeltaE_in_prediction_area.png}
		%		\caption{\vspace{.2cm}}
		%		\label{fig:results:rel_deviations_2d:b}
		%	\end{subfigure}
	%	\begin{subfigure}[t]{0.32\textwidth}
		%		\includegraphics[width=\linewidth]{GRAY_WITHTUMOR_exemplary_data_2d_rat16_with_DeltaE_in_prediction_area.png}
		%		\caption{\vspace{.2cm}}
		%		\label{fig:results:rel_deviations_2d:c}
		%	\end{subfigure}
	\begin{subfigure}[t]{0.35\textwidth}
		\includegraphics[width=\linewidth]{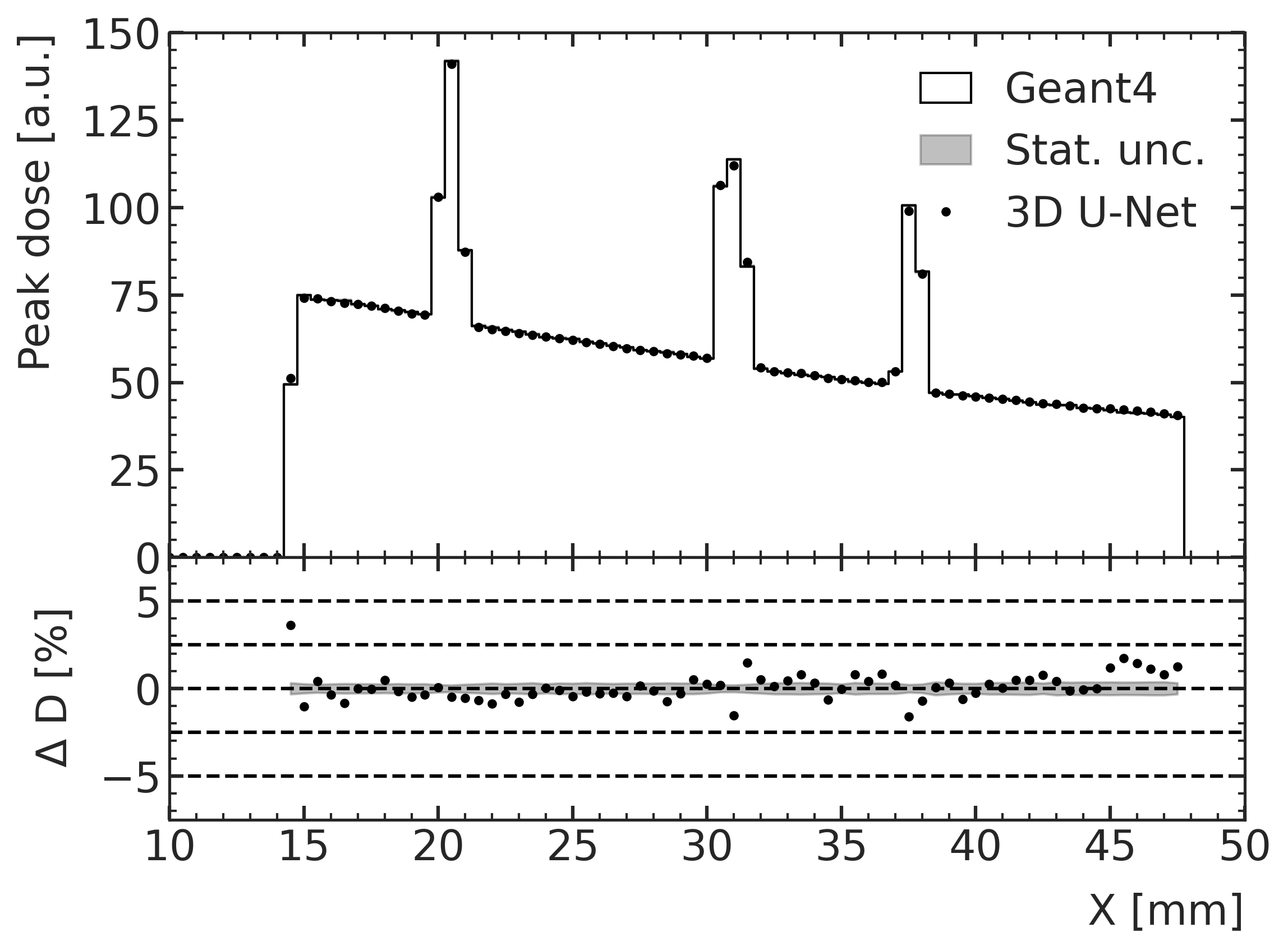}
		\caption{\vspace{.2cm}}
		\label{fig:results:1D_test_comparisons:d}
	\end{subfigure}
	%	\begin{subfigure}[t]{0.32\textwidth}
		%		\includegraphics[width=\linewidth]{GRAY_peak_example_rat_15_which=8_withUnc.png}
		%		\caption{\vspace{.2cm}}
		%%	\end{subfigure}
	%	\begin{subfigure}[t]{0.32\textwidth}
		%		\includegraphics[width=\linewidth]{GRAY_peak_example_rat_16_which=8_withUnc.png}
		%		\caption{\vspace{.2cm}}
		%		\label{fig:results:1D_test_comparisons:f}
		%	\end{subfigure}
	\begin{subfigure}[t]{0.35\textwidth}
		\includegraphics[width=\linewidth]{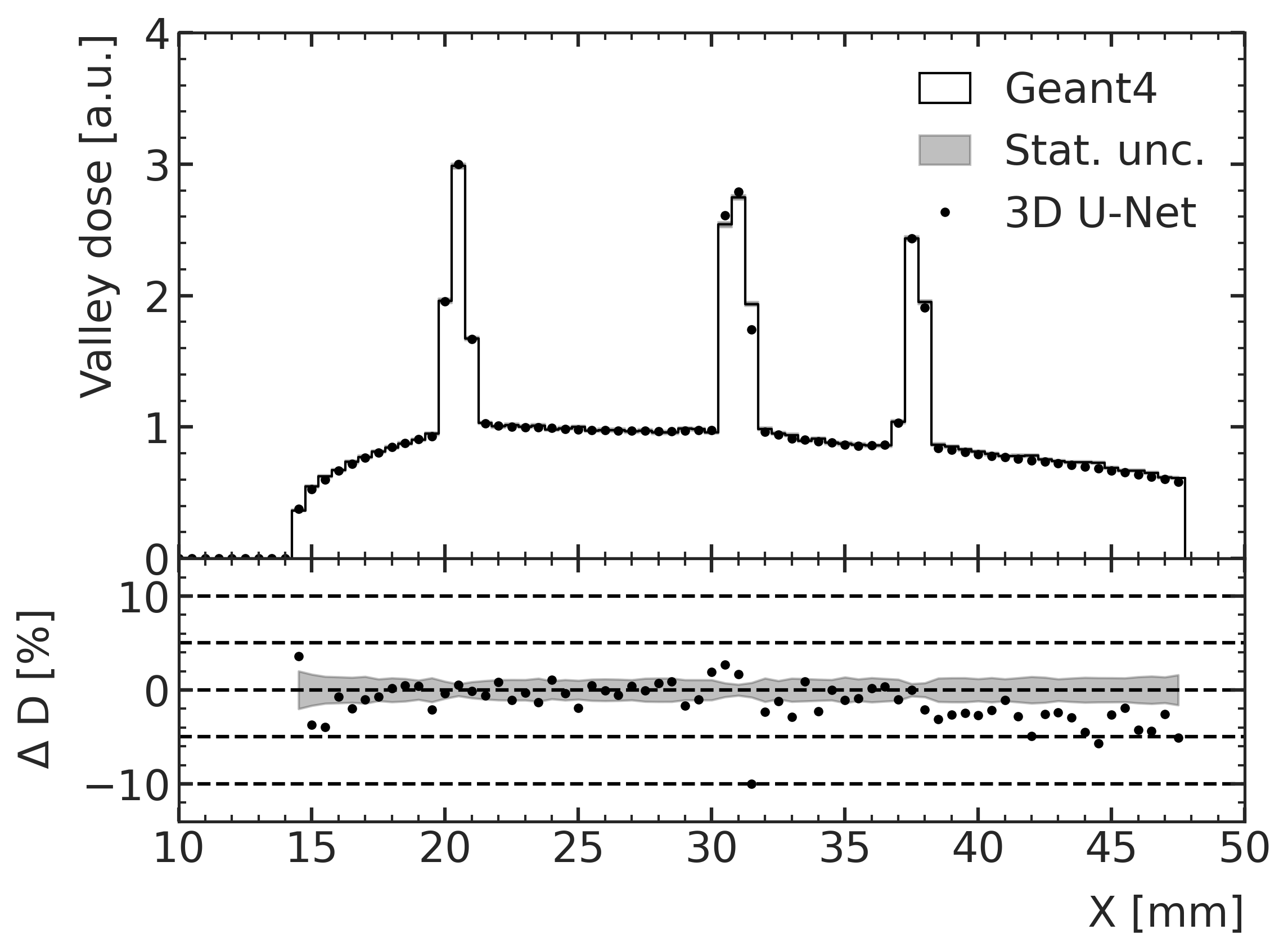}
		\caption{}
		\label{fig:results:1D_test_comparisons:g}
	\end{subfigure}
	%	\begin{subfigure}[t]{0.32\textwidth}
		%		\includegraphics[width=\linewidth]{GRAY_example_rat_15_which=8_withUnc.png}
		%%%%	\begin{subfigure}[t]{0.32\textwidth}
			%	\includegraphics[width=\linewidth]{GRAY_example_rat_16_which=8_withUnc.png}
			%\caption{}
			%\label{fig:results:1D_test_comparisons:i}
			%\end{subfigure}
		\caption{(a) Relative dose difference ($\Delta D_{rel}$) between ML and MC model,  for test rat number 14, in the peak and valley regions. The tumour volume in the shown slice is indicated with a white overlay. (b) depth-peak dose curves;  (c) depth-valley dose curve at the centre of the prediction volume. Doses are normalized to the valley doses at the centre of the brain. }%\textcolor{red}{ Red arrows in the relative dose deviation subplot below indicate deviations larger than the shown ranges.}}
	\label{fig:results:rel_deviations_2d}
\end{figure}

\noindent The energy deposition predictions for the three treatment cases described in Section~\ref{Tumours} are converted to dose in units of Gray to link the results more directly to their preclinical implications. The target geometries around the treated tumours and the respective peak and valley dose prediction deviation are shown in Figure~(\ref{fig:results:rel_deviations_2d}) for an exemplary case (rat number 14, for which the lowest agreement between ML prediction and MC calculation was found) and compared to the low-noise MC simulation data. The figure shows the percentage difference of relative dose and the depth-dose curves for the peak and valley doses at the centre of the prediction volume.  
At least 93.9\% of all voxels of the peak dose prediction and at least 77.6\% of all voxels of the valley dose prediction exhibit less than 3\% dose deviation (see Table~\ref{tab:gamma_treatment}). Especially in the region of the tumour, indicated with a white overlay in Figure~(\ref{fig:results:rel_deviations_2d}), the agreement is very high; a deviation of at most 3\% is achieved for at least 95.9\% of the valley dose voxels and 100.0\% of the peak dose voxels, respectively. Towards the distal end of the phantom, systematic deviations of the ML prediction from the MC simulation can be seen either over- or underestimating the doses, mostly within 10\% agreement, which is the case for 98.5\% and 97.6\% voxels in the peaks and valleys, respectively. 

\noindent The respective fractions of voxels with a difference in terms of relative dose $\Delta D_{rel}<$ 3\% in the full prediction volume, in the tissue volume only and in the tumour volume only are shown in Table~(\ref{tab:gamma_treatment}). Especially the peak predictions show agreements within 3\% for over 93\% of the full prediction volume and 100\% for the tumour targets. The valley dose predictions exhibit a larger fraction of deviating voxels which may be explained by the larger statistical uncertainties of the valley dose MC training data when compared to the peak dose data. Nevertheless, we obtain an agreement within the tumour volume above 95\% also for the valley dose. 

\begin{table}[t]
	\centering
	\renewcommand{\arraystretch}{1.2}
	\begin{center}
		\caption{Fraction of voxels with a relative deviation of dose $\Delta D_{rel}$ between ML prediction and low-noise MC simulation of less than 3\% in the peak and valley regions, shown for the full phantom, only tissue parts of the phantom (the bone voxels have not been considered) and the treated tumour volumes.}\label{tab:gamma_treatment}
		\begin{tabularx}{\textwidth}{>{\raggedright\arraybackslash}X >{\raggedright\arraybackslash}X >{\centering\arraybackslash}X >{\centering\arraybackslash}X >{\centering\arraybackslash}X}	
			\toprule
			Rat ID & Peak/Valley  & \multicolumn{3}{>{\hsize=\dimexpr3\hsize+3\tabcolsep+\arrayrulewidth\relax}X}{Voxel ratio with $\Delta D_{rel}$ < 3\% [\%]}\\
			\midrule
			& & Full phantom & Tissue only & Tumour volume\\
			\midrule
			14 & Peak & 93.9 & 95.0 & 100.0 \\
			& Valley & 77.6 & 81.0 & 95.9 \\
			15 & Peak & 93.9 & 95.7 & 100.0 \\
			& Valley & 81.1 & 85.0 & 97.7 \\
			16 & Peak & 94.6 & 96.1 & 100.0 \\
			& Valley & 80.1 & 83.8 & 97.9 \\
			\bottomrule
		\end{tabularx}
	\end{center}
\end{table}

\begin{figure}[t]
	\centering
	\begin{subfigure}[t]{0.58\textwidth}
		\includegraphics[width=\linewidth]{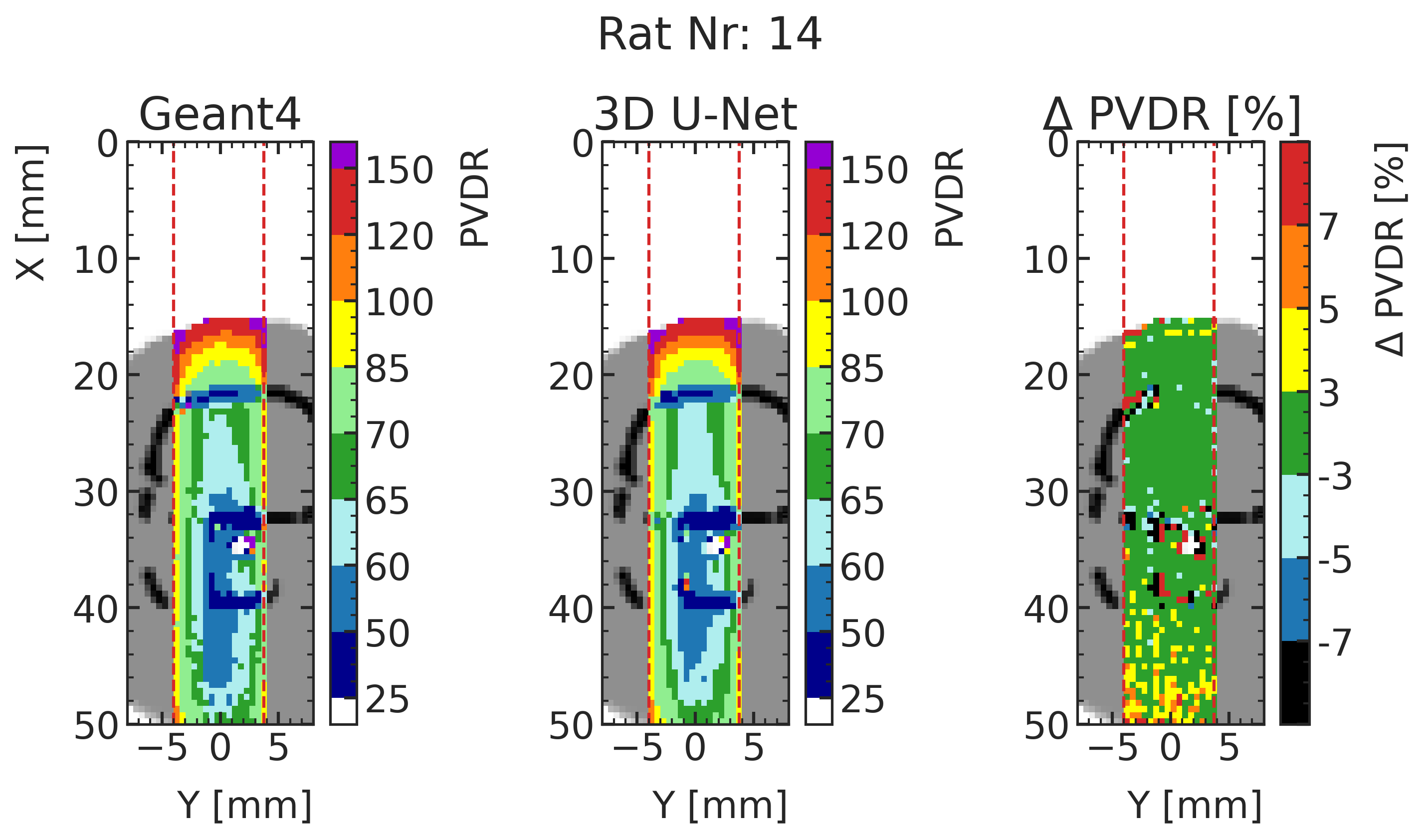}
		\caption{}
		\label{fig:results:PVDR:a}
	\end{subfigure}
	\begin{subfigure}[t]{0.38\textwidth}
		\includegraphics[width=\linewidth]{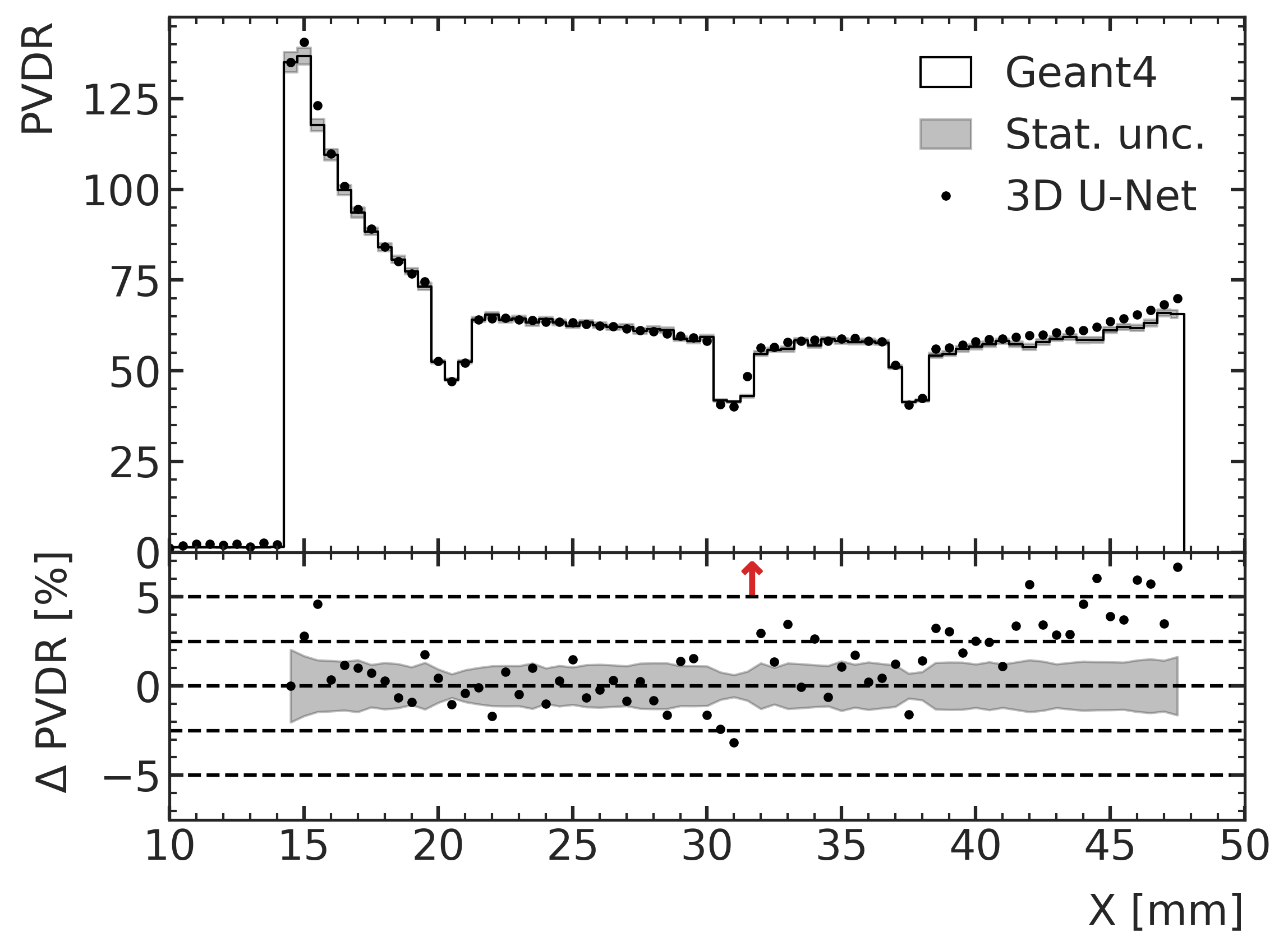}
		\caption{}
		\label{fig:results:PVDR:b}
	\end{subfigure}
	\caption{Exemplary comparison of PVDR computed from MC simulation and ML prediction. $\Delta$ PVDR is calculated as $(PVDR_{ML}-PVDR_{MC})/PVDR_{ML}$ for each macro voxel (see Section~\ref{scoring}). Red arrows in the relative PVDR deviation subplot indicate deviations larger than the shown ranges.}
	\label{fig:results:PVDR}
\end{figure}

\noindent By predicting both the peak and valley doses, the biologically important PVDR can also be calculated with the ML model (see Section~\ref{scoring}). As an example, the predicted PVDR for test rat 14 around the treatment site is shown in Figure~(\ref{fig:results:PVDR}). Comparing the deviations with Figure~(\ref{fig:results:rel_deviations_2d:a}) it can be seen that the deviations of the valley dose predictions are the main driver of PVDR deviations. In all three test cases, the deviation of the predicted PVDR from MC data is less than 5\% for approximately 97\% of all voxels, averaged over the three rats, for peak predictions and approximately 94\% of all voxels for the valley predictions. Figure~(\ref{fig:results:PVDR:b}) shows the values and deviations together with the respective statistical uncertainty along the centre of the prediction volume.

\noindent Using the ML models of the peak and valley regions, it is possible to assess the impact of hypothetical treatment planning using the ML models. During preclinical rat treatment, according to the method defined in ~\cite{Engels2020} and~\cite{Paino}, the irradiation duration is defined by choosing a prescription valley dose $D^*$ and exposing the patient to as much irradiation as needed to achieve a minimum valley dose of $D^*$ in the entire tumour, obtaining 100\% coverage. Using this prescription method, the minimum valley dose predictions using ML and MC are compared to assess the resulting difference in applied dose to the rats. The resulting differences are shown in Table~(\ref{tab:dose_differences}) and are at maximum approximately 1\%. This means that a treatment plan based on the predictions of the ML model would be acceptably accurate in terms of total delivered dose, when compared to MC simulations. 
\begin{table}[t]
	\centering
	\renewcommand{\arraystretch}{1.2}
	\begin{center}
		\caption{Deviation in delivered dose ($\Delta D$) due to a difference in predicted minimum valley dose between ML prediction and MC simulation, in the entire prediction region ($8 \times 8 cm^2$ field size).} \label{tab:dose_differences}
		\begin{tabularx}{\textwidth}{ >{\raggedright\arraybackslash}X
				>{\raggedright\arraybackslash}X >{\raggedright\arraybackslash}X >{\raggedright\arraybackslash}X
			}
			\toprule
			& Rat 14  & Rat 15 & Rat 16 \\
			\midrule
			$\Delta$ D [\%] & 1.17 & 0.95 & -0.37 \\
			\bottomrule
		\end{tabularx}
	\end{center}
\end{table}

\section{Discussion\label{discussion}}

\noindent This study presents an essential step in advancing fast dose prediction models for MRT treatment planning. Although trained on relatively high-noise MC data with mean statistical fluctuations of 5\% (15\%) in the peak (valley) region, the developed ML model exhibits dose deviations of under 3\% compared to low-noise MC simulation data for most voxels (> 95\%) in the tumour volume in the three exemplary treatment cases under study. The prediction of a dose distribution (using a pre-processed density matrix as input) takes approximately 50$\,$ms, which is significantly faster than the currently fastest calculation method which takes approximately 30$\,$minutes~\cite{Donzelli2018}. Batch processing allows the simultaneous prediction (up to 32 samples with an Nvidia GForce 1080t GPU with 11 GB memory). When compared to high-noise MC simulations, the ML dose engine is  approximately one million
times faster to predict the peak and valley doses in the macro-voxels. To note, the MC simulations are executed on CPUs, while the ML dose engine on GPUs.

\noindent While the achieved dose prediction accuracy of more than 3\% within nearly all of the tumour volume for both the peak (100.0\% of voxels) and valley (> 95.5\% of voxels), its dosimetric performance may be improved outside the target tumor, close to air cavities and bone structures, increasing the training set with a larger number of CT scans from rats. The difference of the ML model performance between the peak and the valley dose predictions may be partly explained by the different noise in the used training MC data. Neverthless, the ML model predictions agree very well even when compared to the low-noise MC data. Only around 20\% of the voxels of the full phantom or around 4\% of the voxels in the tumor volume deviate by more than 3\% although the model was trained with on average 5\% and 15\% noisy data for peaks and valleys, respectively. This gives us confidence that the model generalizes well and learns a very good approximate of the underlying function despite the noise. 

\noindent While the presented ML training method and the achieved results provide a very promising outlook for future studies, important limitations of the study need to be stated. 
The ML model is trained only on a single set of beam characteristics, such as energy profile, divergence, and fixed MRT field size. In addition, the used target phantoms exhibit only limited variation as irradiations are performed only from the top of the skull. Another limitation, which probably is common when developing ML dose engines for radiotherapy treatments at the preclinical stage, is the small number of CT scans that are available, therefore it was necessary to augment the data artificially. Another implication arising from the relatively small number of data points, especially the three independent test subjects, is that individual features of each of them contributes strongly to the final results. While this is notable in the reported results, the performances on the training, validation, and test subjects were found to be statistically comparable between those data points. Regarding the presented model trained only on the limited number of CT scans available, we believe that this shows a sufficient degree of generalization. Regarding future studies, the success of the approach should be reproduced and validated with a larger number of test subjects with a larger degree of variation to show the generalization capability of such a model more significantly. 
When comparing the performances on test and training data, small systematic biases occur, which tend to hinder the accurate prediction of the valley doses around air cavities and bones, especially outside the tumor. Still, despite this limitation, the dose prediction of the ML dose engine is satisfactory and, when applied to a clinical environment, the ML dose engine  should generate  even better results than shown here, thanks to the availability of a larger training data set, which would translate in a better predictive power of the ML model. Another limitation is that the ML dose engine needs to be adapted and, at least, partially retrained when applying it to other cases (e.g.~changing target phantom, filters and magnetic field of the MRT beamline, radiotherapy treatment).

\noindent Our results indicate that it is feasible to train ML prediction models with satisfactory accuracy using relatively high-noise MC training data. The clear advantage of using high-noise MC simulations is the  acceleration of the training of the ML dose engine when applied to spatially fractionated therapies such as MRT, for which MC simulations are usually very time-consuming. The high-noise MC samples used in this study in the training and validation are acquired with 1/40th of the simulation time of the low-noise MC simulations. The significantly reduced execution times of the high-noise MC training data actually allows for easy adaptation to different preclinical conditions. Investigating the percentage of voxels exhibiting a dose deviation of less than one standard deviation  and comparing it to the expectation of 68\% allows for a meaningful interpretation of smooth ML predictions. This is more difficult using only measures such as the MAE, which is mainly driven by the deviation between ML prediction and MC simulation caused by the statistical uncertainty of the data. 

\noindent{An aspect for future studies would be the quantitative investigation of the dependency of the dose prediction accuracy of the ML dose engine  on the statistical uncertainty of the training data, which was not investigated in the scope of this study. In this work, the presented high-noise MC simulation data were chosen exemplary.}

\noindent{In future extensions of the developed ML model, a considerable additional benefit could be achieved by increasing the training set and including larger prediction volumes, especially out-of-field, for a better estimation of doses to organs at risk (OAR) around the tumour, which is currently not considered in preclinical treatments at the Australian Synchrotron}.

\section{Conclusion\label{conclusion}}

\noindent This study presents the first successful application of a ML model for MRT dose prediction in a preclinically relevant scenario. The training data comprises high-noise MC data which is much faster to acquire than low-noise MC data. The resulting ML predictions are smooth and do not exhibit the noise present in the MC data. A comparison with low-noise test data shows that the predicted doses are accurate within 3\% for at least 77.6\% of all predicted voxels (at least 95.9\% of voxels containing tumour) in the case of the valley dose prediction and for at least 93.9\% of all predicted voxels (100.0\% of voxels containing tumour) in the case of the peak dose prediction. The ML model seems to generalise well even if we use training MC data with relatively high statistical uncertainty~(15\%).

\noindent The findings of this study allow for an optimistic outlook for the development of ML models to quickly predict doses for preclinical and especially spatially fractionated treatments, which usually require long MC simulation times. Future studies will translate the findings to other MRT treatment settings, including different beam modalities, conformal irradiations and new target phantoms.

\noindent

\section*{Acknowledgements}
\noindent The authors gratefully acknowledge the computing time provided on the Linux HPC cluster at Technical University Dortmund (LiDO3), partially funded in the course of the Large-Scale Equipment Initiative by the German Research Foundation (DFG) as project 271512359.
\\
The authors acknowledge the contribution of the University of Wollongong with NHMRC Near Miss funding. 

\authorcontributions{Contributions are ordered alphabetically. Conceptualization: S. Guatelli, M. Hagenbuchner, M. Lerch,  F. Mentzel, O. Nackenhorst; methodology: F. Mentzel, J. Paino; software: F. Mentzel, J. Paino; investigation: F. Mentzel; resources: M. Barnes, M. Cameron, S. Corde, E. Engels, S. Guatelli, M. Lerch, F. Mentzel, J. Paino, A. Rosenfeld, M. Tehei, S. Vogel; writing---original draft preparation: F. Mentzel; writing---review and editing  M. Barnes, S. Corde, S. Guatelli, M. Hagenbuchner, M. Lerch, O. Nackenhorst, A. C. Tsoi, J. Weingarten; visualization: F. Mentzel; supervision: S. Corde, S. Guatelli, K. Kröninger, M. Lerch, O. Nackenhorst, A. Rosenfeld, M. Tehei, A. C. Tsoi and J. Weingarten; funding acquisition: S. Guatelli, K. Kröninger, M. Lerch, F. Mentzel, A. Rosenfeld, M. Tehei. All authors have read and agreed to the published version of the manuscript. Please turn to the  \href{http://img.mdpi.org/data/contributor-role-instruction.pdf}{CRediT taxonomy} for the term explanation.}
\begin{adjustwidth}{-\extralength}{0cm}
	
	\reftitle{References}
	
	
	\bibliography{references.bib}

\end{adjustwidth}
\end{document}